\documentclass{nature}
\usepackage{epsfig}
\usepackage{multirow}

\spacing{1.0}

\def\MIT{1}
\def\Harvard{2}
\def\KICPChicago{3}
\def\EFIChicago{4}
\def\CfA{5}
\def\UMD{6}
\def\GSFC{7}
\def\UChicago{8}
\def\Munich{9}
\def\ExcellenceCluster{10}
\def\PhysicsUChicago{11}
\def\Miss{12}
\def\AAUChicago{13}
\def\ANL{14}
\def\NIST{15}
\def\PUC{16}
\def\McGill{17}
\def\PennState{18}
\def\Berkeley{19}
\def\UFlorida{20}
\def\Colorado{21}
\def\NASA{22}
\def\Davis{23}
\def\LBNL{24}
\def\Caltech{25}
\def\Arizona{26}
\def\Michigan{27}
\def\MPE{28}
\def\CaseWestern{29}
\def\Minnesota{30}
\def\STScI{31}
\def\SAIC{32}
\def\Yale{33}
\def\BCCP{34}

\usepackage[fleqn]{amsmath}

\newcommand\aj{{Astron.\ J.}}%
\newcommand\araa{{ARA\&A}}%
\newcommand\apj{{Astrophys.\ J.}}%
\newcommand\apjl{{Astrophys.\ J.\ Lett.}}%
\newcommand\apjs{{Astrophys.\ J.\ Supp.}}%
\newcommand\aap{{Astron.\ Astrophys.}}%
\newcommand\mnras{{Mon.\ Not.\ R.\ Astron.\ Soc.}}%
\newcommand\pasp{{PASP}}%
\newcommand{\etal}{et~al.~}

\title{A Massive, Cooling-Flow-Induced Starburst in the Core of a Highly Luminous Galaxy Cluster}

\author{M.~McDonald$^{\MIT}$, 
M.~Bayliss$^{\Harvard}$, 
B.~A.~Benson$^{\KICPChicago,\EFIChicago}$, 
R.~J.~Foley$^{\CfA}$, 
J.~Ruel$^{\Harvard}$, 
P.~Sullivan$^{\MIT}$, 
S.~Veilleux$^{\UMD,\GSFC}$, 
K.~A.~Aird$^{\UChicago}$, 
M.~L.~N.~Ashby$^{\CfA}$, 
M.~Bautz$^{\MIT}$, 
G.~Bazin$^{\Munich,\ExcellenceCluster}$, 
L.~E.~Bleem$^{\KICPChicago,\PhysicsUChicago}$, 
M.~Brodwin$^{\Miss}$, \\
J.~E.~Carlstrom$^{\KICPChicago,\PhysicsUChicago,\EFIChicago,\AAUChicago,\ANL}$, 
C.~L.~Chang$^{\KICPChicago,\EFIChicago,\ANL}$, 
H.~M. Cho$^{\NIST}$, 
A.~Clocchiatti$^{\PUC}$, 
T.~M.~Crawford$^{\KICPChicago,\AAUChicago}$, 
A.~T.~Crites$^{\KICPChicago,\AAUChicago}$, 
T.~de~Haan$^{\McGill}$, 
S.~Desai$^{\Munich,\ExcellenceCluster}$, 
M.~A.~Dobbs$^{\McGill}$, 
J.~P.~Dudley$^{\McGill}$, 
E.~Egami$^{\Arizona}$,
W.~R.~Forman$^{\CfA}$, 
G.~P.~Garmire$^{\PennState}$, 
E.~M.~George$^{\Berkeley}$, 
M.~D.~Gladders$^{\KICPChicago,\AAUChicago}$, 
A.~H.~Gonzalez$^{\UFlorida}$, 
N.~W.~Halverson$^{\Colorado}$, 
N.~L.~Harrington$^{\Berkeley}$, 
F.~W.~High$^{\KICPChicago,\AAUChicago}$, 
G.~P.~Holder$^{\McGill}$, 
W.~L.~Holzapfel$^{\Berkeley}$, 
S.~Hoover$^{\KICPChicago,\EFIChicago}$, 
J.~D.~Hrubes$^{\UChicago}$, 
C.~Jones$^{\CfA}$, 
M.~Joy$^{\NASA}$, 
R.~Keisler$^{\KICPChicago,\PhysicsUChicago}$, 
L.~Knox$^{\Davis}$, 
A.~T.~Lee$^{\Berkeley,\LBNL}$, 
E.~M.~Leitch$^{\KICPChicago,\AAUChicago}$, 
J.~Liu$^{\Munich,\ExcellenceCluster}$, 
M.~Lueker$^{\Berkeley,\Caltech}$, 
D.~Luong-Van$^{\UChicago}$, 
A.~Mantz$^{\KICPChicago}$, 
D.~P.~Marrone$^{\Arizona}$, 
J.~J.~McMahon$^{\KICPChicago,\EFIChicago,\Michigan}$, 
J.~Mehl$^{\KICPChicago,\AAUChicago}$, 
S.~S.~Meyer$^{\KICPChicago,\PhysicsUChicago,\EFIChicago,\AAUChicago}$, 
E.~D.~Miller$^{\MIT}$,
L.~Mocanu$^{\KICPChicago,\AAUChicago}$, 
J.~J.~Mohr$^{\Munich,\ExcellenceCluster,\MPE}$, 
T.~E.~Montroy$^{\CaseWestern}$, 
S.~S.~Murray$^{\CfA}$, 
T.~Natoli,$^{\KICPChicago,\PhysicsUChicago}$, 
S.~Padin$^{\KICPChicago,\AAUChicago,\Caltech}$, 
T.~Plagge$^{\KICPChicago,\AAUChicago}$, 
C.~Pryke$^{\Minnesota}$, 
T.~D.~Rawle$^{\Arizona}$,
C.~L.~Reichardt$^{\Berkeley}$, 
A.~Rest$^{\STScI}$, 
M.~Rex$^{\Arizona}$,
J.~E.~Ruhl$^{\CaseWestern}$, 
B.~R.~Saliwanchik$^{\CaseWestern}$, 
A.~Saro$^{\Munich,\ExcellenceCluster}$, 
J.~T.~Sayre$^{\CaseWestern}$, 
K.~K.~Schaffer$^{\KICPChicago,\EFIChicago,\SAIC}$, 
L.~Shaw$^{\McGill,\Yale}$, 
E.~Shirokoff$^{\Berkeley,\Caltech}$, 
R.~Simcoe$^{\MIT}$, 
J.~Song$^{\Michigan}$, 
H.~G.~Spieler$^{\LBNL}$, 
B.~Stalder$^{\CfA}$, 
Z.~Staniszewski$^{\CaseWestern}$, 
A.~A.~Stark$^{\CfA}$, 
K.~Story$^{\KICPChicago,\PhysicsUChicago}$, 
C.~W.~Stubbs$^{\CfA,\Harvard}$, 
R.~\v{S}uhada$^{\Munich}$, 
A.~van~Engelen$^{\McGill}$, 
K.~Vanderlinde$^{\McGill}$, 
J.~D.~Vieira$^{\KICPChicago,\PhysicsUChicago,\Caltech}$, 
A. Vikhlinin$^{\CfA}$, 
R.~Williamson$^{\KICPChicago,\AAUChicago}$, 
O.~Zahn$^{\Berkeley,\BCCP}$, 
and
A.~Zenteno$^{\Munich,\ExcellenceCluster}$
}

\begin{document}
\maketitle

\begin{affiliations}
\item MIT Kavli Institute for Astrophysics and Space Research, Massachusetts Institute of Technology, 77 Massachusetts Avenue, Cambridge, MA 02139
\item Department of Physics, Harvard University, 17 Oxford Street, Cambridge, MA 02138
\item Kavli Institute for Cosmological Physics, University of Chicago, 5640 South Ellis Avenue, Chicago, IL 60637
\item Enrico Fermi Institute, University of Chicago, 5640 South Ellis Avenue, Chicago, IL 60637
\item Harvard-Smithsonian Center for Astrophysics, 60 Garden Street, Cambridge, MA 02138
\item Department of Astronomy, University of Maryland, College Park, MD 20742
\item Astroparticle Physics Laboratory, NASA Goddard Space Flight Center, Code 661, Greenbelt, MD 20771 USA
\item University of Chicago, 5640 South Ellis Avenue, Chicago, IL 60637
\item Department of Physics, Ludwig-Maximilians-Universit\"{at, Scheinerstr.\ 1, 81679 M\"{u}nchen, Germany}
\item Excellence Cluster Universe, Boltzmannstr.\ 2, 85748 Garching, Germany
\item Department of Physics, University of Chicago, 5640 South Ellis Avenue, Chicago, IL 60637
\item Department of Physics and Astronomy, University of Missouri, 5110 Rockhill Road, Kansas City, MO 64110
\item Department of Astronomy and Astrophysics, University of Chicago, 5640 South Ellis Avenue, Chicago, IL 60637
\item Argonne National Laboratory, 9700 S. Cass Avenue, Argonne, IL, USA 60439
\item NIST Quantum Devices Group, 325 Broadway Mailcode 817.03, Boulder, CO, USA 80305
\item Departamento de Astronomia y Astrosifica, Pontificia Universidad Catolica, Chile
\item Department of Physics, McGill University, 3600 Rue University, Montreal, Quebec H3A 2T8, Canada
\item Department of Astronomy and Astrophysics, Pennsylvania State University, 525 Davey Laboratory, University Park, PA 16802
\item Department of Physics, University of California, Berkeley, CA 94720
\item Department of Astronomy, University of Florida, Gainesville, FL 32611
\item Department of Astrophysical and Planetary Sciences and Department of Physics, University of Colorado, Boulder, CO 80309
\item Department of Space Science, VP62, NASA Marshall Space Flight Center, Huntsville, AL 35812
\item Department of Physics,  University of California, One Shields Avenue, Davis, CA 95616
\item Physics Division, Lawrence Berkeley National Laboratory, Berkeley, CA 94720
\item California Institute of Technology, 1200 E. California Blvd., Pasadena, CA 91125
\item Steward Observatory, University of Arizona, 933 North Cherry Avenue, Tucson, AZ 85721
\item Department of Physics, University of Michigan, 450 Church Street, Ann   Arbor, MI, 48109
\item Max-Planck-Institut f\"{ur extraterrestrische Physik, Giessenbachstr.\ 85748 Garching, Germany}
\item Physics Department, Center for Education and Research in Cosmology  and Astrophysics,  Case Western Reserve University, Cleveland, OH 44106
\item Physics Department, University of Minnesota, 116 Church Street S.E., Minneapolis, MN 55455
\item Space Telescope Science Institute, 3700 San Martin Dr., Baltimore, MD 21218
\item Liberal Arts Department, School of the Art Institute of Chicago,  112 S Michigan Ave, Chicago, IL 60603
\item Department of Physics, Yale University, P.O. Box 208210, New Haven, CT 06520-8120
\item Berkeley Center for Cosmological Physics, Department of Physics, University of California, and Lawrence Berkeley
National Labs, Berkeley, CA 94720

\end{affiliations}

\begin{abstract} 
In the cores of some galaxy clusters the hot intracluster plasma is dense enough that it should cool radiatively in the cluster's lifetime\cite{lea73,cowie77, fabian77}, leading to continuous ``cooling flows'' of gas sinking towards the cluster center, yet no such cooling flow has been observed. The low observed star formation rates\cite{odea08,mcdonald11b} and cool gas masses\cite{edge01} for these ``cool core'' clusters suggest that much of the cooling must be offset by astrophysical feedback to prevent the formation of a runaway cooling flow\cite{mcnamara07,fabian12,mathews09,gomez02}. 
Here we report X-ray, optical, and infrared observations of the galaxy cluster SPT-CLJ2344-4243\cite{williamson11} at $z$ = 0.596.
These observations reveal an exceptionally luminous ($L_{2-10~{\rm keV}} = 8.2\times10^{45}$ erg s$^{-1}$) galaxy cluster which hosts an extremely strong cooling flow (\.{M}$_{cool}$ = 3820 $\pm$ 530 M$_{\odot}$ yr$^{-1}$).
Further, the central galaxy in this cluster appears to be experiencing a massive starburst (740 $\pm$ 160 M$_{\odot}$ yr$^{-1}$), which suggests that the feedback source responsible for preventing runaway cooling in nearby cool core clusters may not yet be fully established in SPT-CLJ2344-4243. 
This large star formation rate implies that a significant fraction of the stars in the central galaxy of this cluster may form via accretion of the intracluster medium, rather than the current picture of central galaxies assembling entirely via mergers. 
\end{abstract}

\begin{figure*}[p]
\begin{center}
\includegraphics[width=0.95\textwidth]{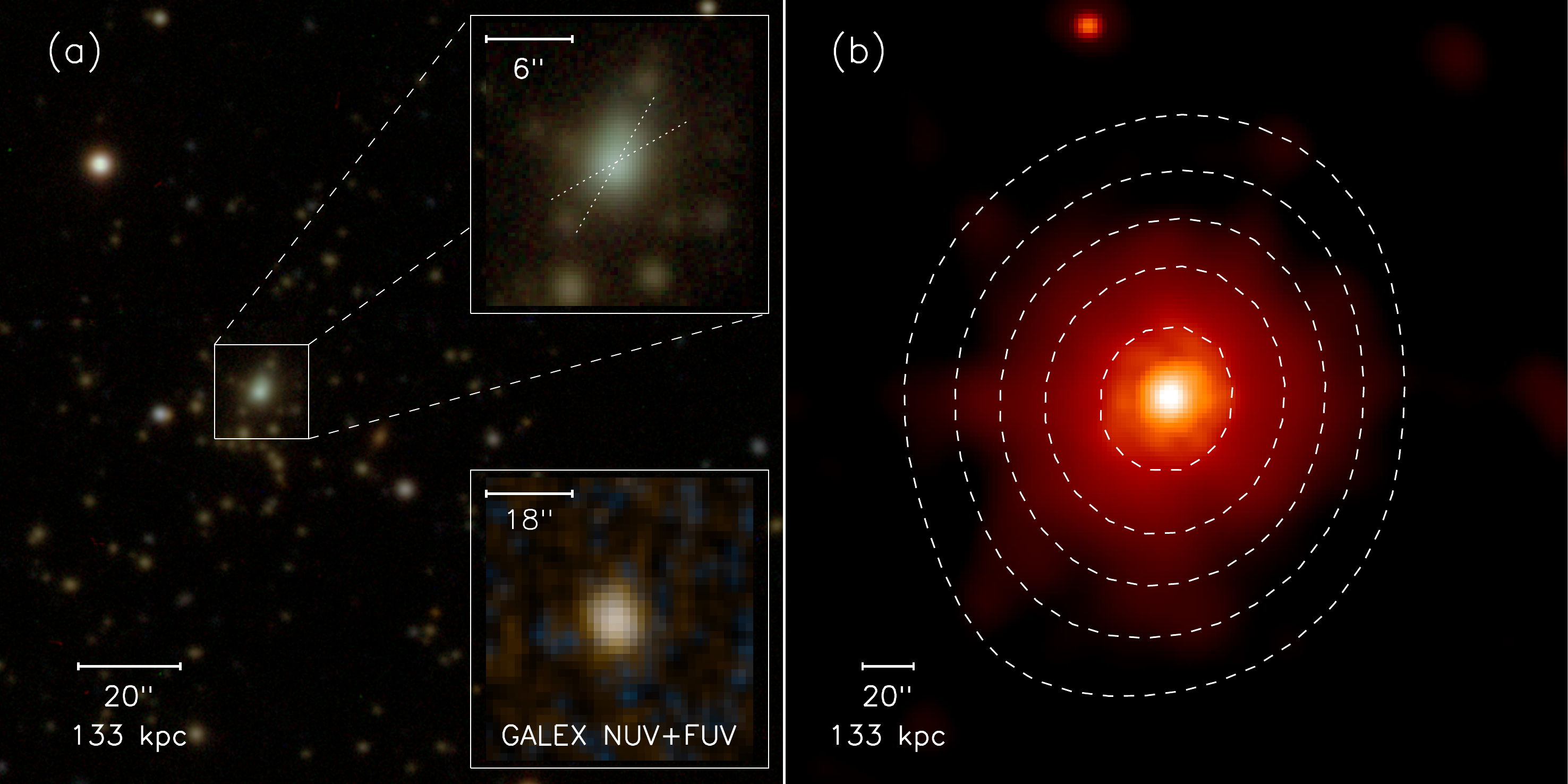}
\caption{False-color images of the galaxies and intracluster plasma in the galaxy cluster SPT-CLJ2344-4243. (a): This color-composite image of SPT-CLJ2344-4243 is based on an RGB combination of optical $r$, $i$, $z$ images. Galaxies which make up the galaxy cluster share a common brown color, due to their similar star formation histories and common distance. The central galaxy, which is both the most massive and most luminous galaxy in the cluster, is considerably bluer than the rest of the member galaxies, suggesting significantly younger stellar populations. This is more obvious in the zoomed-in inset. The lower right inset, which shows an ultraviolet color-composite, reveals a bright UV source, with no accompanying emission from the surrounding member galaxies. Dotted lines represent the orientation of the optical and near-IR long-slit spectra. (b): This false-color image shows the adaptively-smoothed X-ray data, with photon energies from 0.7--2.0 keV (to minimize AGN contribution), of SPT-CLJ2344-4243. This image clearly shows the luminous, centrally-concentrated core, as well as the relatively smooth, relaxed morphology of the intracluster medium. White contours represent the SZ decrement (significance levels of 5, 10, 15, 20, 25) against the cosmic microwave background. The circularity of these contours agree with the scenario that this system is not currently undergoing a a major merger with another galaxy cluster.}
\label{broadband}
\end{center}
\end{figure*}

The galaxy cluster SPT-CLJ2344-4243 was discovered by the South Pole Telescope (hereafter SPT\cite{carlstrom11}) via the Sunyaev-Zel'dovich (SZ) effect, with an initial estimated mass of $M_{200}$ $\sim$16.6 $\times$ 10$^{14}$ M$_{\odot}$\cite{williamson11}. These data were supplemented with new broadband optical $g, r, i, z$ imaging from the Mosaic~II camera on the Blanco 4-m telescope (Figure \ref{broadband}), optical multi-object spectroscopy using GMOS on the Gemini South 8.1-m telescope, optical long-slit spectroscopy using IMACS on the 6.5-m Magellan telescope, near-infrared long-slit spectroscopy using FIRE on the 6.5-m Magellan telescope, mid--far infrared imaging using PACS and SPIRE on the Herschel Space Observatory, and X-ray imaging spectroscopy using the ACIS-I camera on the Chandra X-ray Observatory. Additionally, we have acquired archival near--far UV imaging from the GALEX archives, near--mid infrared imaging from the 2MASS and WISE archives, and 843MHz radio imaging from the SUMSS survey. Further details for these data and their processing can be found in Supplementary Information.

We estimate the mass of SPT-CLJ2344-4243 from the X-ray-measured pressure ($Y_X$ $\equiv$ $M_{gas}$ $\times$ $T_X$) of the intracluster medium (ICM), using an externally calibrated pressure-mass ($Y_X$--$M$) relation.  The relation was calibrated using a local sample of relaxed clusters from X-ray estimates of the total mass that assumed hydrostatic equilibrium\cite{vikhlinin09a}. 
By iteratively adjusting the value of $r_{500}$ (where $r_{500}$ (or $r_{200}$) is the radius for which the enclosed average density is 500 (200) times the critical (average) density of the Universe) such that the $Y_X$--$M_{500}$ relation is satisfied, we converge on values of $r_{500}$ = 1.3 Mpc and $M_{500, Y_X}$ = 12.6$^{+2.0}_{-1.5}$ $\times$10$^{14}$ M$_{\odot}$. At r$_{200}$, this corresponds to $M_{200, Y_X}$ $\sim$ 25 $\times$10$^{14}$ M$_{\odot}$, which makes SPT-CLJ2344-4243 amongst the most massive known clusters in the Universe\cite{menanteau11,foley11}. The GMOS multi-object spectroscopy of 26 galaxies exhibiting only absorption features yielded a robust biweight estimate of the redshift ($z$ = 0.596 $\pm$ 0.002) and velocity dispersion (1700$^{+300}_{-200}$~ km s$^{-1}$), the latter being consistent  with the picture of an extremely massive cluster. The velocity distribution is consistent with a Gaussian distribution, but the limited number of redshifts does not preclude velocity substructure or multimodality.
The smooth X-ray isophotes suggest that the cluster may be relaxed,
and while the cluster member velocity distribution is consistent with
an undisturbed cluster, the velocity data lack the statistical power
to robustly constrain the cluster's dynamical state.

The integrated rest-frame 2--10 keV X-ray luminosity, $L_{2-10~{\rm keV}}$ = 8.2$^{+0.1}_{-0.2}$ $\times$ 10$^{45}$ erg s$^{-1}$ within $r_{500}$, is greater than any other known cluster in this band. The high central luminosity, which is predominantly cooling radiation, in turn results in a high X-ray cooling rate, as defined by $\frac{dM}{dt}$ = $\frac{2L\mu m_p}{5kT}$, where $\mu$ is the mean molecular mass of the ICM. Assuming a cooling radius of 100~kpc (see Supplementary Information), we measure an ICM cooling rate of 3820 $\pm$ 530 M$_{\odot}$ yr$^{-1}$, making this the strongest cooling flow yet discovered (see Table 1 for comparison to other clusters). The ICM in SPT-CLJ2344-4243 exhibits a significant drop in temperature, accompanied by a rise in the metallicity, in the central 100 kpc, reminiscent of nearby cool core clusters. Furthermore, the short central cooling time ($<$ 1 Gyr), along with the low central entropy  ($<$100 keV cm$^2$), resembles nearby ``strong'' cool cores, such as the Perseus\cite{fabian00} and PKS0745-191\cite{allen96} clusters. The discovery of a strong cool core at $z$ = 0.596 is particularly remarkable as recent X-ray and optical surveys have found a general lack of strong cool cores at $z$ $>$ 0.4\cite{vikhlinin07,santos08,mcdonald11c}, with relatively few exceptions.

\begin{figure}[p]
\centering
\includegraphics[height=0.6\textheight]{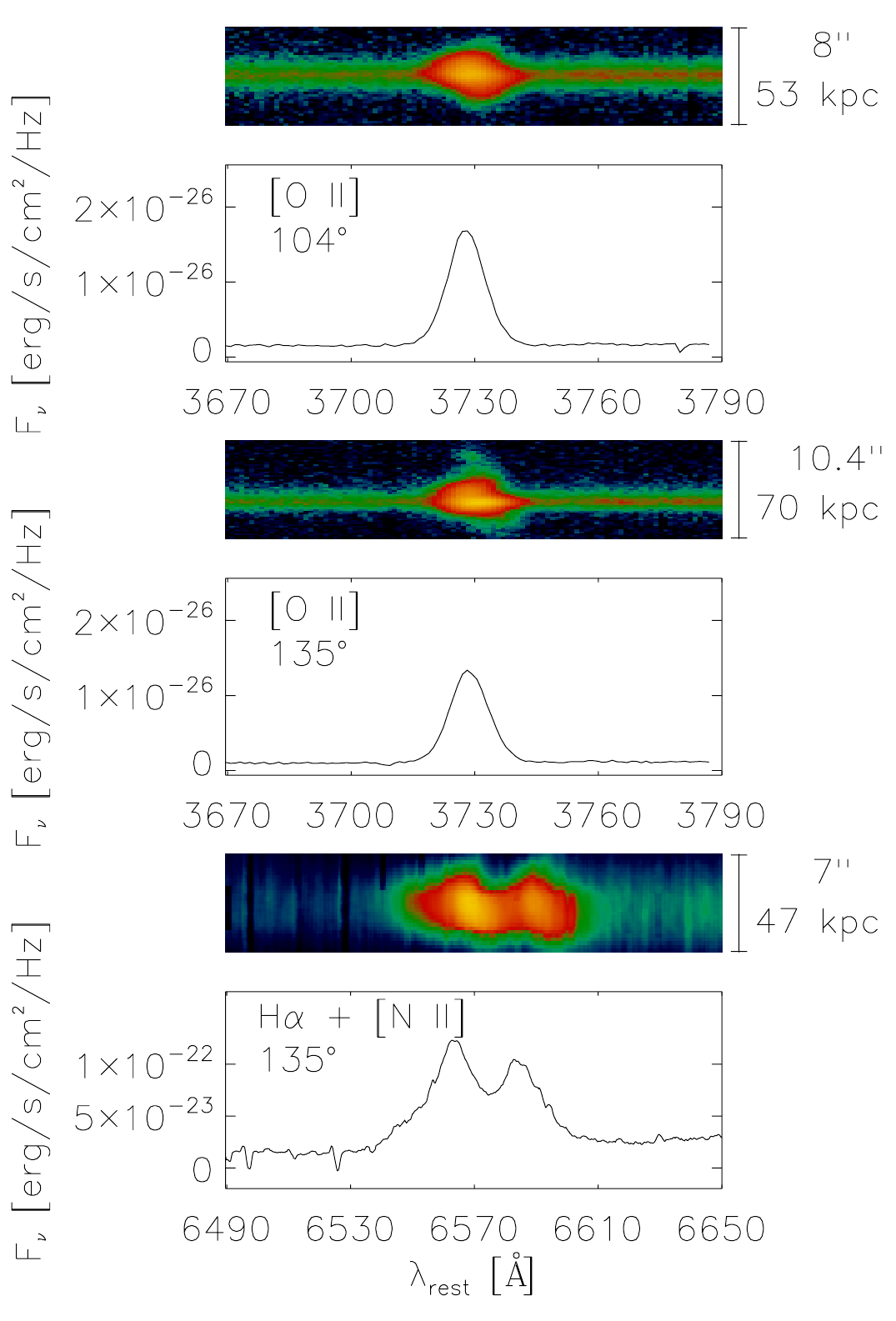}
\caption{Optical and near-infrared emission-line spectra of the central galaxy in SPT-CLJ2344-4243. The false-color images highlight the extended line emission ([O II] in upper two panels, H$\alpha$+[N II] in lower), where the vertical axis is the spatial direction (along the slit) and the horizontal axis is the spectral direction. These emission lines result from warm, ionized gas at $\sim$10$^4$ K, which is likely heated by a combination of young stars, shocks, and feedback from the central active black hole. The fact that the [O II] emission is significantly more extended in one direction ($\theta$ $\sim$ 135$^{\circ}$) suggests that the emission is non-axisymmetric, and is consistent with the scenario of radial line-emitting filaments. The extent of this emission ($>$ 50 kpc) is reminiscent of optical filaments observed in the core of the Perseus cluster\cite{conselice01}. Beneath each color image, we show the spectrum which is generated by summing along columns of the color image. These spectra show the high signal-to-noise of these emission lines, leading to high-confidence estimates of the emission line luminosity. The lower panel, which shows the near-infrared spectrum (in the observed frame, $\lambda_{H\alpha}$ = 1.05$\mu$m) shows emission from both the H$\alpha$ and [N II] lines, extended over similar radii ($>$ 50 kpc). 
}
\label{imacsspec}
\end{figure}

Much like the central galaxies of low-$z$ cool core clusters\cite{hu85,heckman89,mcdonald10}, SPT-CLJ2344-4243 exhibits bright, spatially-extended, optical line emission (i.e. [O II], H$\beta$, [O III], [O I], H$\alpha$, [N II], [S II], etc; Figure \ref{imacsspec}). We were fortunate to intersect what appears to be an extended filament with one of our randomly-oriented slits, which has a length of $\sim$70 kpc. This is similar in extent to the most extended optical filaments in the core of the Perseus cluster\cite{conselice01}, and orders of magnitude larger than typical jets in cluster cores (e.g., M87\cite{sparks96}). The diagnostic line ratios [N II]/H$\alpha$, [S II]/H$\alpha$, [O I]/H$\alpha$, [O III]/H$\beta$, and [O III]/[O II] show evidence for a Seyfert-like AGN in the nucleus of the central galaxy, while at radii $>$2$^{\prime\prime}$, the optical line ratios resemble those in the star-forming filaments of $z$~$\sim$~0 cool core clusters\cite{mcdonald12a} (see Supplementary Information).
 
\begin{figure}[p]
\centering
\includegraphics[height=0.6\textheight]{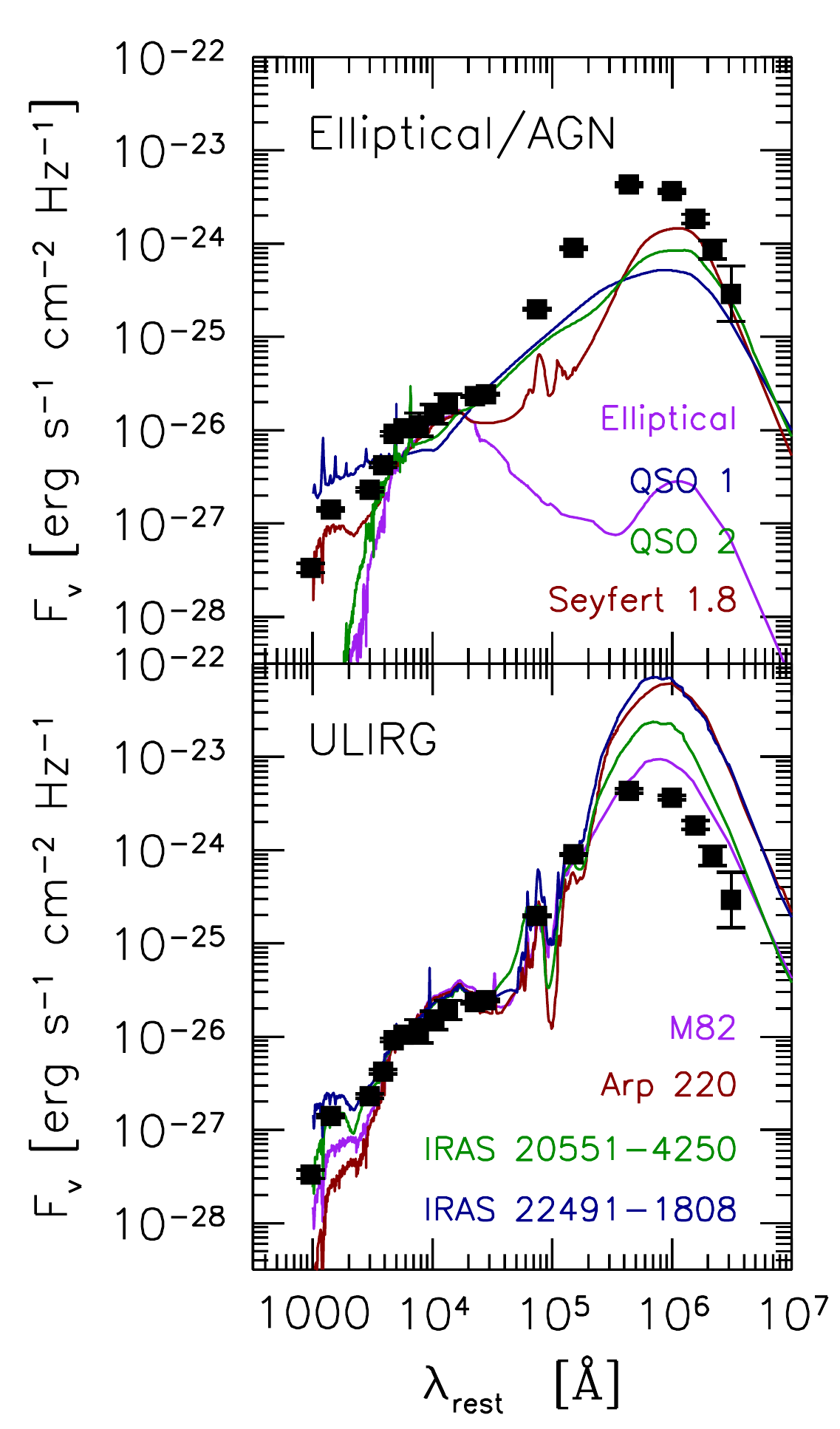}
\caption{Spectral energy distribution of the central galaxy in SPT-CLJ2344-4243 from the far-UV to the far-infrared. We show on the Y-axis the integrated specific flux (F$_{\nu}$) as a function of wavelength in the rest frame of the galaxy cluster, along with the associated 1$\sigma$ measurement error. A typical central cluster galaxy is morphologically-classified as an ``elliptical'' galaxy, and has an average spectral energy distribution shown in purple in the upper panel. In contrast, the central galaxy in SPT-CLJ2344-4243 has a considerable excess of emission at both ultraviolet and infrared wavelengths, indicative of strong star formation. While there is significant evidence for strong feedback from the central, supermassive black hole in this cluster (see Supplementary Information), this figure shows that simple models of active galactic nuclei (quasar type 1 and 2, and Seyfert type 1.8 shown in upper panel) are unable to reproduce the spectral shape of the central galaxy in SPT-CLJ2344-4243. However, in the lower panel we show templates\cite{polletta07} of four different dusty starbursts, or ultra-luminous infrared galaxies (ULIRGS), which provide a much better match to the data. This good agreement suggests that the central galaxy in SPT-CLJ2344-4243, unlike typical central cluster galaxies, contains a dusty starburst and heavily obscured AGN. Specifically, the spectral shape is most similar to those of M82, a dusty starburst with a strong wind, and IRAS 20551-4250, which is a composite of a highly-obscured AGN and a vigorous, dusty starburst. 
}
\label{sed}
\end{figure}

Apart from the exceptionally X-ray high luminosity and central cooling rate, what truly sets this system apart from the majority of nearby galaxy clusters is that there is significant evidence for a dusty starburst in the central galaxy of SPT-CLJ2344-4243. The rest-frame 0.1--500$\mu$m spectral energy distribution (Figure \ref{sed}) of the central galaxy most closely resembles that of an ultraluminous, infrared galaxy (ULIRG), which are known for having heavily obscured starbursts ($\sim$200--1000 M$_{\odot}$ yr$^{-1}$) and central AGN. This scenario is corroborated by our observation of significant Balmer reddening ($E(B-V)_{global}$ $\sim$ 0.3) and strong 24--160$\mu$m emission, combined with signatures of ongoing star formation (near- and far-UV emission, bright nebular emission lines, weak 4000\AA\ break) and a heavily-obscured central AGN ($E(B-V)_{nuclear}$ $\sim$ 0.5, $n_{H,X-ray}$ $\sim$ 40 $\times$ 10$^{22}$ cm$^{-2}$). Utilizing the full multi-wavelength dataset, which includes X-ray, near--far UV, optical, near--far IR, and radio data, we estimate an extinction-corrected, AGN-subtracted star formation rate of 740 $\pm$ 160 M$_{\odot}$ yr$^{-1}$, assuming a geometric correction of 45\% for the long-slit spectroscopy and an AGN contamination fraction of $\sim$40-50\% (see Supplementary Information for details). 
 
The presence of extended ($\sim$70 kpc), morphologically-complex (Figure \ref{imacsspec}) star-forming filaments coincident with the central galaxy in SPT-CLJ2344-4243 is reminiscent of low-$z$ cool core clusters like Perseus and PKS0745-191. However, while these clusters have substantial amounts of star formation ($\sim$1--20 M$_{\odot}$ yr$^{-1}$)\cite{odea08,mcdonald11b}, this is still orders of magnitude less than predicted by the classical cooling estimates based on the X-ray luminosity (Table 1). This disagreement has become known as the ``cooling flow'' problem, and it is generally assumed that some form of feedback is responsible for halting the cooling ICM before it reaches the cold phase. 
SPT-CLJ2344-4243, however, represents an exception to this general trend, of which there are very few\cite{mcnamara06, odea08}, where the high star formation rate represents a significant fraction of the massive cooling flow (Table 1).
Whatever feedback mechanism is responsible for preventing runaway cooling of the ICM in low-redshift galaxy clusters is clearly  operating with a lower efficiency in this system. While the central galaxy hosts an active galactic nucleus (AGN), as evidenced by a hard X-ray point source and strong radio emission (see Supplementary Information), it may be that we are observing this system during a small window in time when the AGN is rapidly feeding off of the cooling flow, but the power output of the AGN has not yet fully coupled to the ICM, and therefore is able to halt a smaller fraction of the total cooling than in typical low-redshift clusters (i.e., Perseus).
The fact that systems with such high cooling and star formation rates are not observed at $z=0$ suggests that either this system is entirely unique, or the mechanism which quenches cooling may have been less effective in the early Universe. Further studies of distant, strongly-cooling galaxy clusters are needed to differentiate between these two scenarios.

The high star formation rates inferred from optical line emission and near--far-UV, optical, and mid--far IR continuum emission, combined with the strong signatures of X-ray cooling, suggest that the central galaxy in SPT-CLJ2344-4243 may form a substantial fraction of its stars through an intense, short-lived cooling phase of the intracluster medium. 
Such strong cooling can not be sustained for a significant amount of time, or both the central galaxy and its supermassive black hole would become too massive, and the central galaxy would have stellar populations considerably younger than those observed in giant elliptical galaxies today. 
This implies that episodes of strong cooling are short-lived, in contrast to the longer episodes of strong feedback observed in nearby clusters. 




\begin{addendum}

\item[Acknowledgements] 

M.~McDonald was supported at MIT by NASA through the Chandra X-ray Observatory.
The SPT is supported by the National Science Foundation, with partial support provided by the Kavli Foundation, and the Moore Foundation. Support for X-ray analysis was provided by NASA.  Work at McGill is supported by NSERC, the CRC program, and CIfAR, and at Harvard by NSF. S.~Veilleux acknowledges a Senior NPP Award held at the NASA Goddard Space Flight Center. R.\ Keisler acknowledges a NASA Hubble Fellowship, B.A.\ Benson a KICP Fellowship, M.\ Dobbs an Alfred P. Sloan Research Fellowship, and O.\ Zahn a BCCP fellowship.

\item[Author Contributions] 
M.~McDonald reduced the X-ray and optical long slit spectroscopic data, performed the main analysis, and wrote the paper, with significant assistance from B. Benson, R. Foley, and S. Veilleux, and comments from all other authors. M.~Bayliss and J.~Ruel reduced multi-slit observations of SPT-CLJ2344-4243 and performed the velocity dispersion analysis. P.~Sullivan and R.~Simcoe obtained the infrared spectroscopy, and P.~Sullivan reduced these data. All other authors (listed alphabetically) have contributed as part of the South Pole Telescope collaboration, by either their involvement with the initial cluster discovery in the SZ and/or multi-wavelength follow-up.

\item[Author Information] Reprints and permissions information is available at www.nature.com/reprints. The authors declare that they have no competing financial interests. Correspondence and requests for materials should be addressed to M. McDonald (email: mcdonald@space.mit.edu).

\end{addendum}


\clearpage

\begin{table*}[p]
\begin{center}
{\small
\begin{tabular}{c c c c c c c c}
\hline\hline
Cluster & z & L$_{2.0-10.0 keV}$ & kT & dM/dt & SFR & $\epsilon_{cool}$\\
 & & [10$^{44}$ erg/s] & [keV] & [M$_{\odot}$/yr] & [M$_{\odot}$/yr] & \\
 \hline
 
 Perseus & 0.0179& 11 & 5.5 & 556 & 37 & 0.07\\  

 PKS0745-191 & 0.1028 & 29.5 & 6.71 & 1455 & 20 & 0.01\\ 
 Zw 3146 & 0.2906 & 36.9 & 6.4 & 2228 & 79 & 0.04\\ 
 RX J1347.5-1145 & 0.451 & 60 & 10.0 & 1900 & 23 & 0.01\\ 
 \\
 SPT-CLJ2344-4243 & 0.596 & 82$^{+1}_{-2}$ & 13.0$^{+2.4}_{-3.4}$ & 3820 $\pm$ 530 & 740 $\pm$ 160 & 0.19 $\pm$ 0.05\\
\hline
 \end{tabular}
 \caption{Properties of well-studied, strong cool core clusters, for comparison to SPT-CLJ2344-4243. Previous to this work, RX~J1347.5-1145 was considered both the most X-ray luminous and strongest cooling galaxy cluster, with a luminosity of L$_{2-10\rm{keV}}$ = 60 $\times$ 10$^{44}$ erg s$^{-1}$ and cooling rate of 1900 M$_{\odot}$ yr$^{-1}$. Immediately obvious from this table is the exceptionally high star formation rate of the central galaxy in SPT-CLJ2344-4243. We quantify the efficiency of converting the cooling flow into stars with the parameter $\epsilon_{cool}$, which is simply the star formation rate normalized to the classical cooling rate. The high star formation rate implies that SPT-CLJ2344-4243 is converting $\sim$20\% of the cooling flow into stars, which is considerably higher than the vast majority of low-redshift cool core clusters. 
X-ray properties and star formation rates of the lower-redshift clusters are taken from the literature\cite{allen00,edge01,gitti04}. }
 \label{}
 }
 \end{center}
 \end{table*}
 
\clearpage
\noindent{\large{\bf Supplementary Information}}
\section{Observations, Data Reduction, and Analysis}
\setcounter{figure}{0}
\renewcommand*\thefigure{S.\arabic{figure}}
\setcounter{table}{0}
\renewcommand*\thetable{S.\arabic{table}}
\subsection{X-ray Data:}
Our data reduction pipeline was adapted from earlier work\cite{vikhlinin05, A11} and includes the removal of flares, estimation of background from blank sky fields, and calibration using the latest set of corrections. From the cleaned image, an X-ray spectrum was extracted over the energy range 0.5--8.0 keV within an annulus defined as 0.15r$_{500}$ $<$ r $<$ r$_{500}$, excluding point sources, with the initial assumption that r$_{500}$ = 1.0 Mpc. We fit this spectrum using a combined WABS(MEKAL) model, which accounts for Galactic absorption in the emission spectrum of a hot, diffuse gas. This fit yields a temperature (T$_X$) and gas mass (M$_g$), which are combined to give the mass proxy Y$_X$ $\equiv$ M$_g$ $\times$ T$_X$.  Assuming the following scaling relation\cite{vikhlinin09a}:
\begin{equation}
M_{500} = 5.77\times10^{14} h^{1/2} M_{\odot} \times \left(\frac{Y_X}{3\times10^{14} M_{\odot} keV}\right)^{0.57} E(z)^{-2/5}~,
\end{equation}
we can then infer M$_{500}$, which leads to a new value of r$_{500}$ based on its definition: 
\begin{equation}
r_{500} \equiv \left(\frac{3M_{500}}{4\pi500\rho_{crit}(z)}\right)^{1/3} .
\end{equation}
This process is iterated until changes in $r_{500}$ are small. From the final estimate of M$_{500}$, we can extrapolate M$_{200}$ (which, by convention, uses the average density of the Universe, rather than the critical density) assuming that the dark matter halo has an NFW profile with a concentration following the mass-concentration relation\cite{duffy08}. At this point, we measure T$_X$ within the aperture 0.15r$_{500}$ $<$ r $<$ r$_{500}$, and L$_X$ and M$_g$ within the aperture 0 $<$ r $<$ r$_{500}$, which are used as the final global quantities.

\begin{figure*}[t]
\begin{center}
\begin{tabular}{cc}
\includegraphics[width=0.45\textwidth]{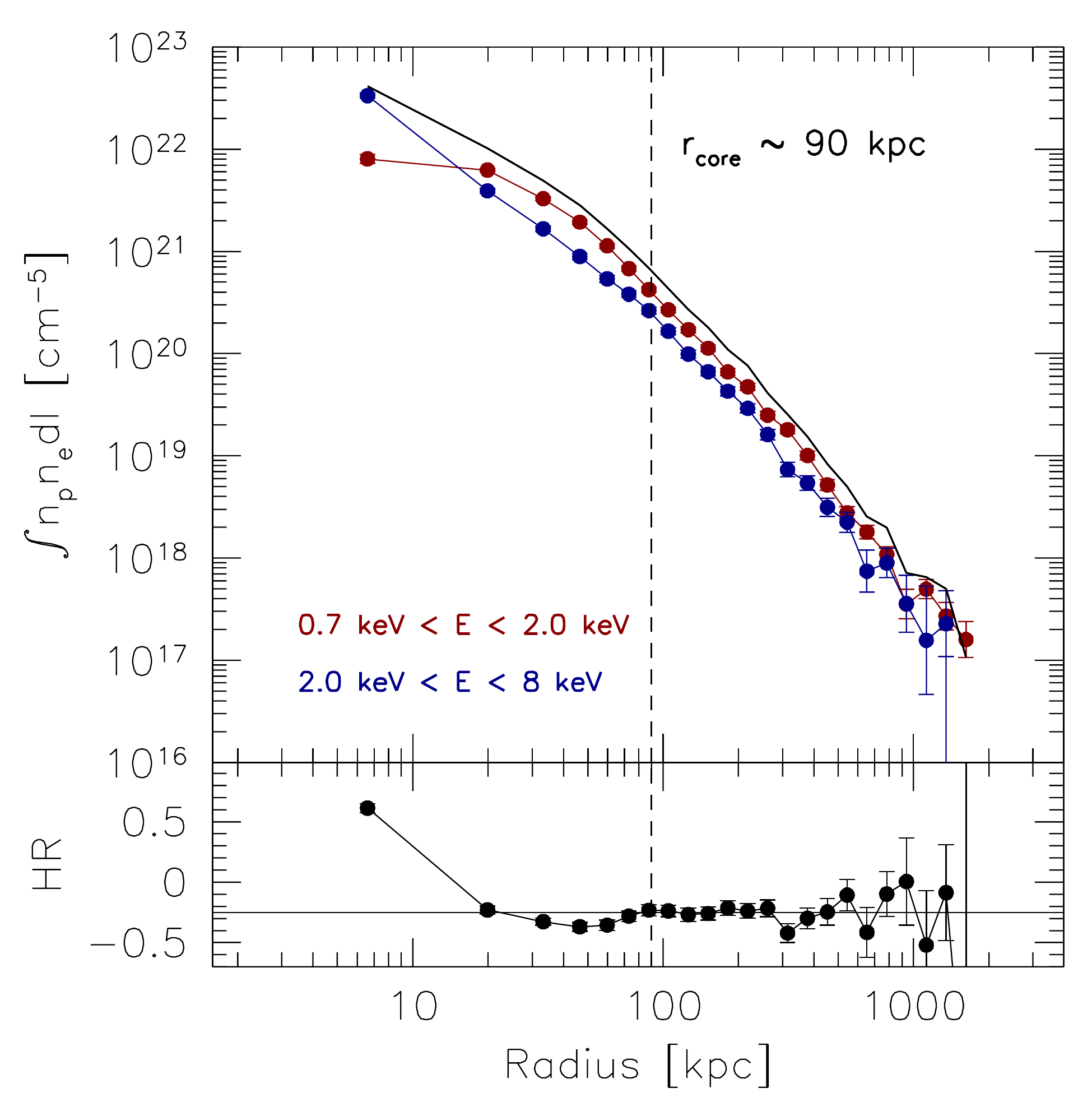} & 
\includegraphics[width=0.48\textwidth]{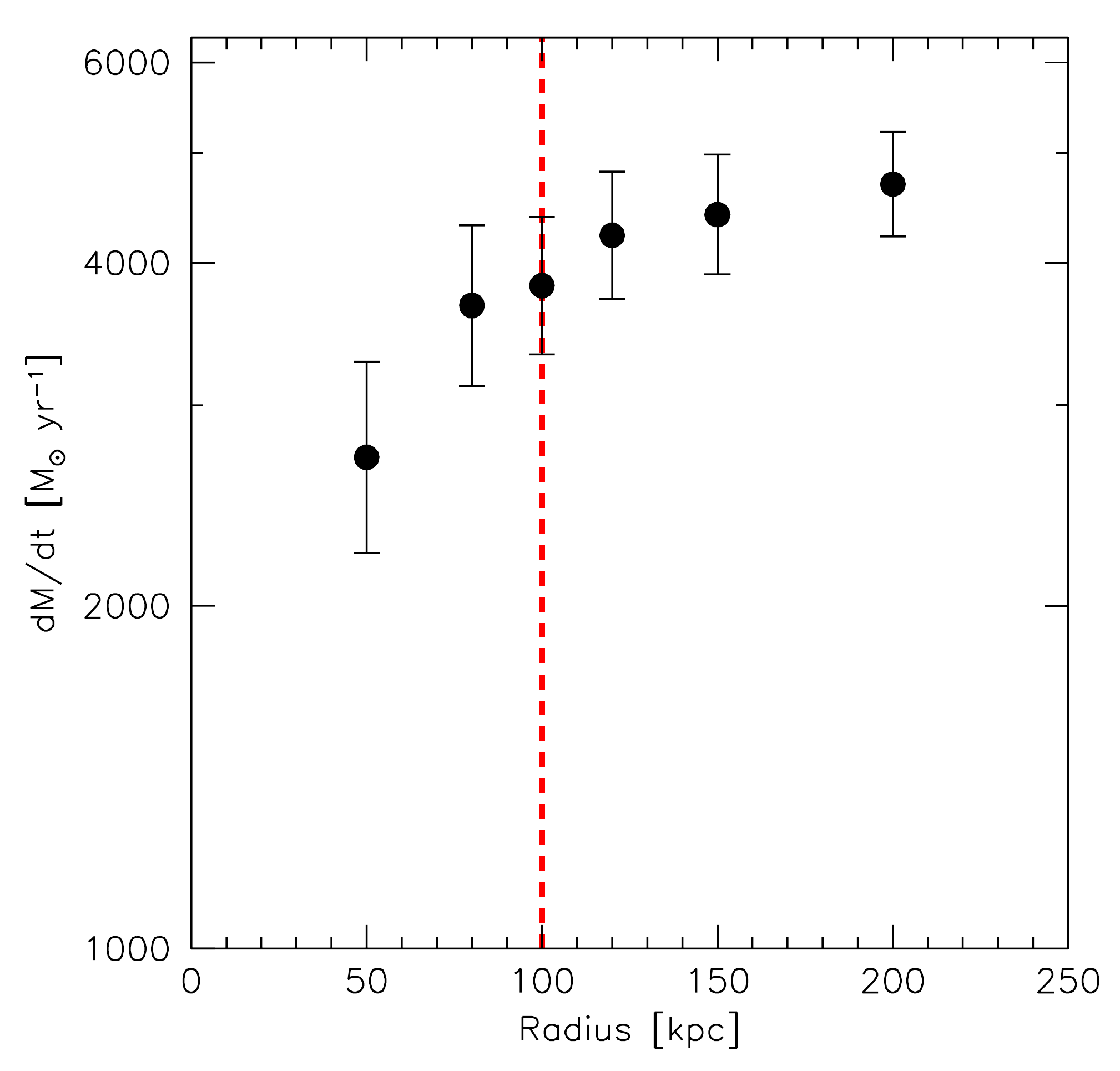} \\
\end{tabular}
\caption{Left: X-ray emission measure as a function of radius for SPT-CLJ2344-4243. The red and blue lines show the contributions to the total (black) from soft (0.7--2.0 keV) and hard (2.0--8.0 keV) emission in the observed frame, respectively. The lower panel shows the hardness ratio, HR = $\frac{H-S}{H+S}$ as a function of radius. 
The spectrally hard, spatially unresolved AGN emission dominates at r $<$ 10 kpc, while the emission from 10 kpc $<$ r $<$ 100 kpc has a soft X-ray excess, which we interpret as a cool core. Note that the error bars are smaller than the point size in this radial range. Right: Classical cooling rate as a function of enclosed radius. This plot shows that the cooling rate rises rapidly out to $\sim$100kpc, at which point it changes little out to $>$200kpc, due to the fact that gas at large radii has a much longer cooling time. This plot further motivates our choice of a 100kpc cooling radius.}
\label{sbprof}
\end{center}
\end{figure*}

To determine an inner characteristic radius, we compute the hard (2.0--8.0 keV) and soft (0.7--2.0 keV) X-ray surface brightness profiles, as shown in Figure \ref{sbprof}. From this plot, it is clear that the central $\sim$10 kpc is dominated by a hard X-ray point source, likely an AGN, and then the hardness ratio reaches a minimum over the range 10--100 kpc, before settling to a roughly constant value from 100--1300 kpc. This inner region with excess soft X-ray emission defines the cool core, with a cooling radius of $\sim$100 kpc. Further, the right panel of Figure \ref{sbprof} shows that the classical cooling rate does not increase substantially by including emission from r $>$ 100 kpc, suggesting that the cool core is confined to r $<$ 100kpc.

We extracted spectra in the logarithmically-spaced annuli 0 $<$ r $<$ 100 kpc (cool core), 100 kpc $<$ r $<$ 450 kpc, and 450 kpc $<$ r $<$ 1300 kpc  in order to determine the gas temperature (T$_X$), electron density (n$_e$), and metallicity (Z) as a function of radius. In the central bin, we masked the inner 1.5$^{\prime\prime}$ in order to remove contributions to the spectrum from the AGN. These spectra were fit with a model combining Galactic absorption (WABS) and hot, diffuse gas (MEKAL).  The specific entropy (K = T$_X$ $\times$ n$_e^{-2/3}$) and cooling time (t$_{cool}$ = 10$^8$ $\left(\frac{K^{3/2}}{10}\right)$$\left(\frac{T_X}{5}\right)^{-1}$ Gyr) were also inferred from the model fit to the spectrum in each annulus.

\begin{figure*}[p]
\begin{center}
\includegraphics[width=0.6\textwidth]{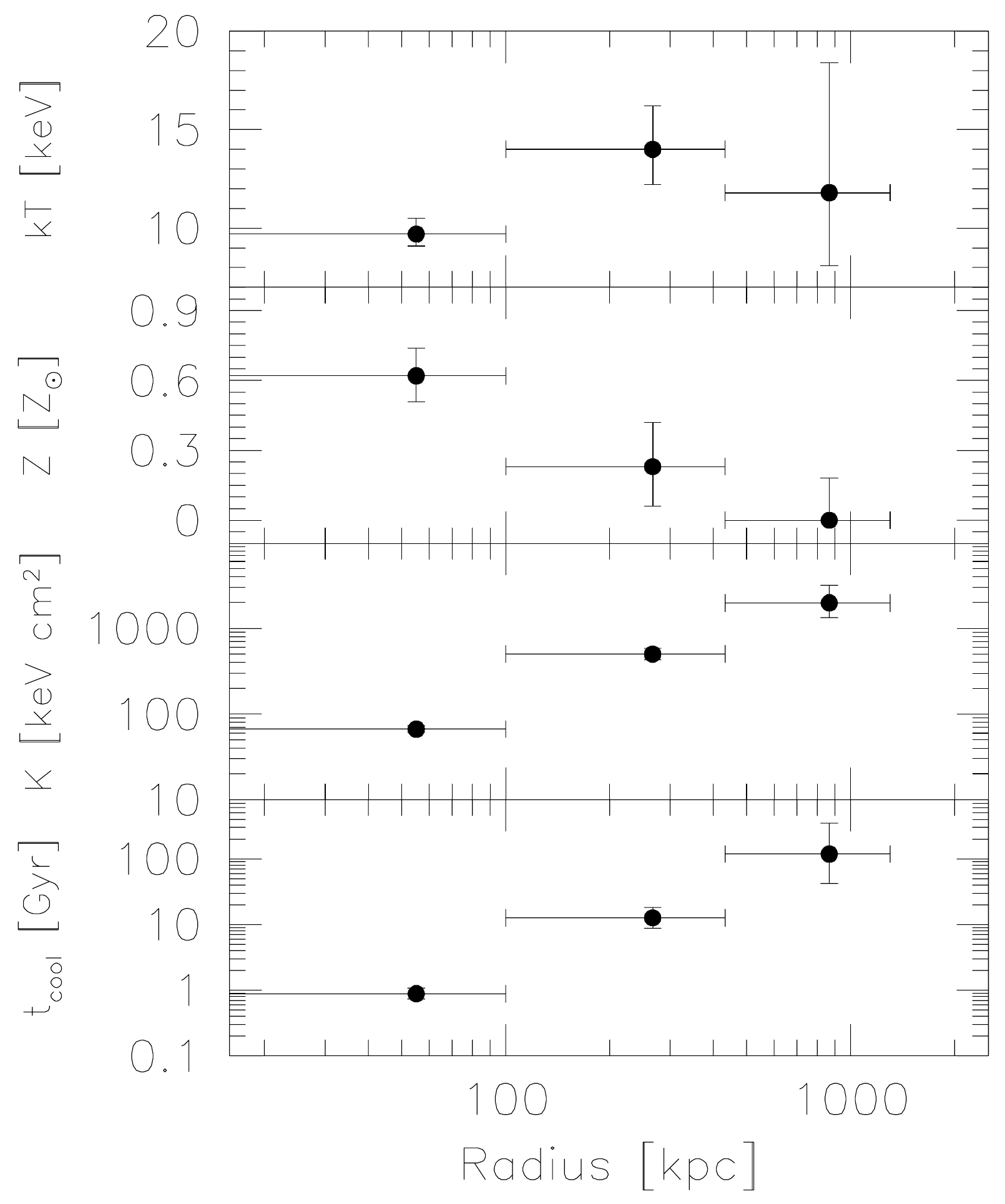}
\caption{Temperature, metallicity, specific entropy, and cooling time profiles for SPT-CLJ2344-4243. The outer radius (1300 kpc) corresponds to $r_{500}$ for this system, while the inner radius (100 kpc) corresponds to the size of the cool core. The dip in temperature and peak in metallicity in the central bin is reminiscent of $z\sim0$ cool core clusters. In the central 100 kpc, we find that, without correcting for projection, the cooling time is $<$ 1 Gyr, suggesting a strong cooling flow.}
\label{xrayprofs}
\end{center}
\end{figure*}

The projected temperature and metallicity profiles show a rise in metallicity accompanied by a dip in temperature in the central 100 kpc, as seen in cool core clusters at z $\sim$ 0. Despite the fact that these data have not been corrected for projection, which means that a significant amount of emission from hot gas at large radius is contributing to the spectrum extracted from the central aperture, the cooling time in the inner 100 kpc is $<$ 1 Gyr. This short central cooling time, along with the low ($<$100 keV cm$^2$) central entropy, resembles nearby strong cool cores, such as the Perseus and PKS0745-191 clusters. 

The classical cooling rate was estimated from the equation, $\frac{dM}{dt}$ = $\frac{2L\mu m_p}{5kT}$, using the total X-ray luminosity and temperature within the central 100 kpc\cite{odea08}. As shown in Figure \ref{sbprof}, this estimate is relatively insensitive to the choice of radius for r $\geq$ 100 kpc. We find dM/dt$_{classical}$ = 3820 $\pm$ 530 M$_{\odot}$ yr$^{-1}$. The available X-ray spectra are of insufficient depth to estimate the cooling rate spectroscopically and thus we are unable to measure the ``instantaneous'' cooling rate. However, the classical, steady-state, cooling rate is still valuable for comparison to low-$z$ clusters and as an upper limit to the ``true'' cooling rate.

\subsection{Infrared Imaging}
Snapshot Herschel PACS images were obtained at 70 and 160$\mu$m with a total exposure time of $<$30 minutes as part of a Director's Discretionary Time project (PI: M. Bayliss). In these short exposures, the central galaxy was detected at flux levels of 432 $\pm$ 21 mJy and 364 $\pm$ 18 mJy, at 70 and 160$\mu$m, respectively.

Herschel SPIRE maps at 250, 350 and 500$\mu$m were observed as part of the ``Herschel Lensing Survey'' (HLS; PI: E. Egami) snapshot program covering 148 SPT clusters. The SPIRE data consists of a single repetition map, with coverage complete to a cluster-centric radius of ~5 arcmin. The maps were produced via the standard reduction pipeline HIPE v9.0, the SPIRE Photometer Interactive Analysis (SPIA) package v1.7, and the calibration product v8.1, with improved treatment of the baseline removal (also known as `de-striping').

To characterize the far-IR SED, we fit a blackbody law, modified with a
spectral emissivity that varies physically such that the dust opacity reaches
unity at frequency $\nu_{\rm c}$ \cite{blain03}:
\begin{equation}
f_{\nu}\propto [1-{\rm exp}(-(\nu / \nu_{\rm c})^\beta)]B_{\nu}(T_{\rm d})
\end{equation}
Here, $B_{\nu}(T_{\rm d})$ is the Planck function. We fix the spectral index of the emissivity to
$\beta = 2.0$, and the critical frequency to $\nu_{\rm c} = 1.5$ THz.  The dust temperature $T_d$ and the amplitude are left as free parameters. We exclude photometric data at wavelengths shorter than
rest wavelength $\sim 40\,{\rm \mu m}$ (including 70, 160, 250, 350, 500 $\mu$m data from Herschel PACS and SPIRE) so we can fit only to the cold dust component, which should be more free of AGN contamination and better trace the dust heated by star formation. Our best fit gives $T_d=87 \pm 3 $\ K and $L_{IR}=(9.5 \pm 1.1)\times 10^{12} ~L_{\odot}$, with $r\chi ^2=0.13$.

\subsection{Optical Spectroscopy:}
\begin{figure*}[t]
\begin{center}
\includegraphics[width=0.6\textwidth]{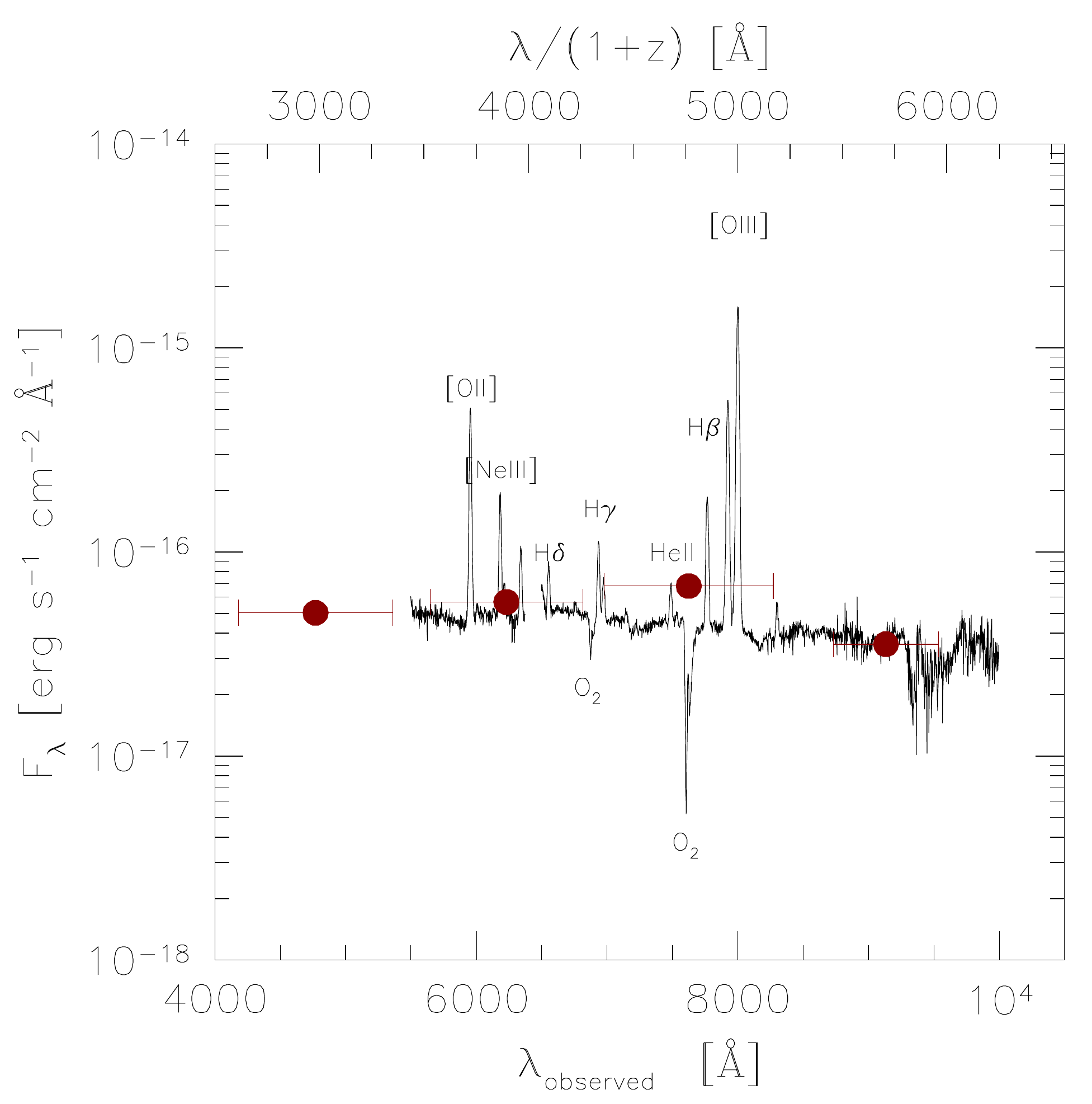}
\caption{IMACS spectrum of the central galaxy in SPT-CLJ2344-4243. This spectrum demonstrates the strength of the nebular emission lines ([O II], H$\beta$, [O III]), as well as the relative flatness of the continuum spectrum around the 4000\AA\ break, indicating a relatively young population of stars. The red circles represent broadband fluxes, extracted in the same region is the spectrum, which were used to flux-calibrate the spectra. }
\label{optsed}
\end{center}
\end{figure*}

Long-slit optical spectra at two different orientations (104$^{\circ}$ and 135$^{\circ}$) with a 1.2$^{\prime\prime}$ slit width were obtained using the IMACS spectrograph on the Baade 6.5m telescope, using the 200 lines/mm grism, which provides 2.0\AA/pixel spectral resolution over the wavelength range 3900\AA--10000\AA. The seeing during these observations was $\sim$0.7$^{\prime\prime}$. These spectra were reduced using standard IRAF tasks to remove the bias and overscan, flat field, remove sky lines, and perform wavelength calibration based on arc lamp spectra. The LA Cosmic software\cite{lacosmic} was used to mask cosmic rays before combining exposures. Flux calibration was performed by measuring the $g$, $r$, $i$, $z$ flux within a 1.2$^{\prime\prime}$$\times$1.2$^{\prime\prime}$ box centered on the central galaxy nucleus from the broadband imaging, and forcing the spectrum, extracted from a boxcar with the same spatial and spectral dimensions, to pass through these points. The final reduced spectrum in a 1.2$^{\prime\prime}$$\times$1.2$^{\prime\prime}$ extraction region is shown in Figure \ref{optsed}.

\begin{figure*}[t]
\begin{center}
\includegraphics[width=0.6\textwidth]{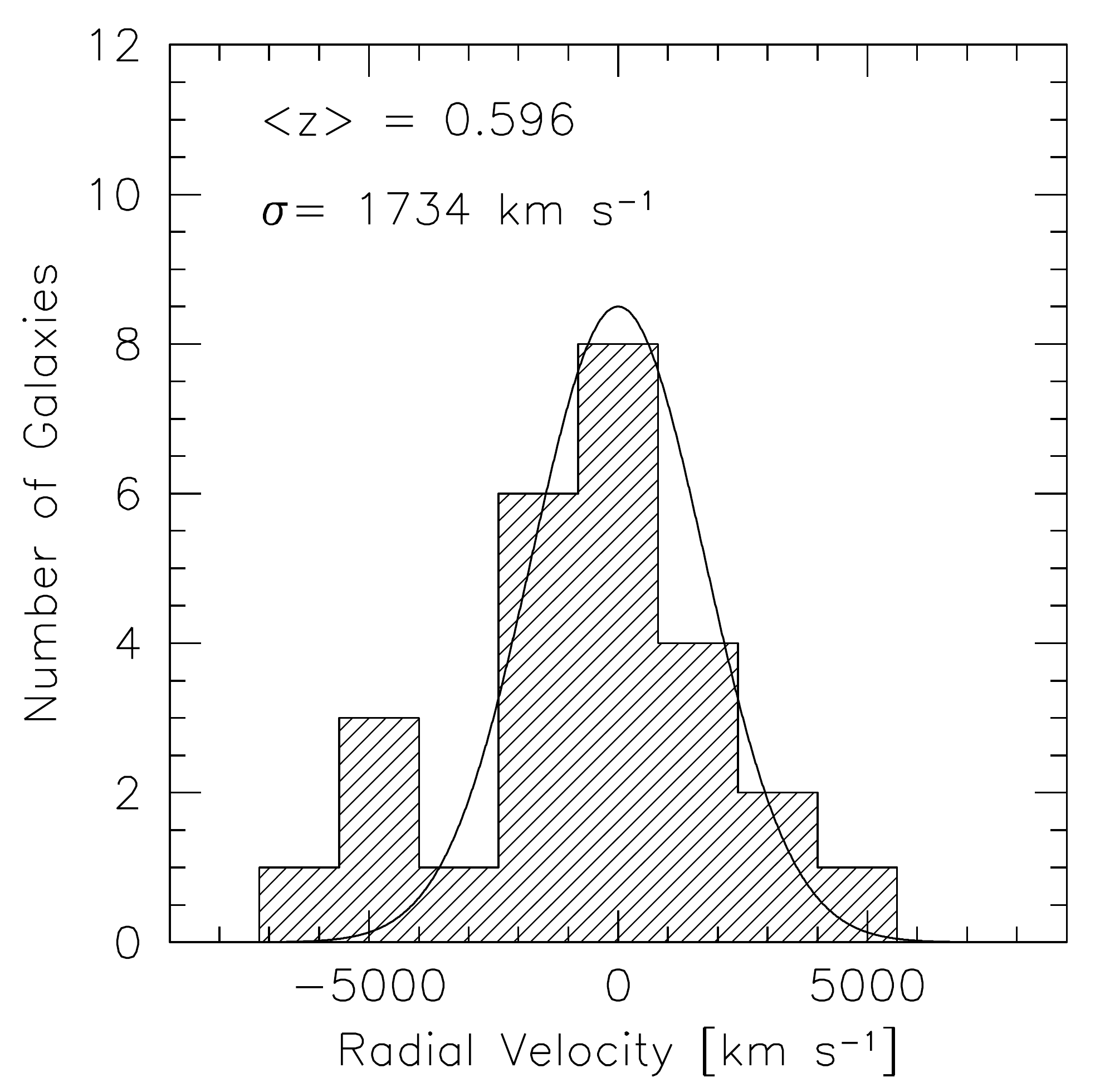}
\caption{Radial velocities of 26 galaxies in SPT-CLJ2344-4243, relative to the mean cluster redshift of $<z>$ = 0.596. The best-fit Gaussian is shown as a solid line. There is some evidence for substructure at negative velocities, but the limited number of redshifts do not allow for a statistically significant result.}
\label{velhist}
\end{center}
\end{figure*}

The velocity dispersion for SPT-CLJ2344-4243 is estimated from cluster member galaxy recession velocities that were measured using the Gemini Multi-Object Spectrograph (GMOS) as a part of a large NOAO survey program (PI: C. Stubbs) to measure velocity dispersions for 100 SPT galaxy clusters. Galaxies were prioritized for MOS slits based on proximity to the red-sequence and magnitude. The raw spectra were bias-subtracted, flat-fielded, wavelength calibrated, and mapped to a common mosaic grid using the \emph{gemini.gmos} IRAF package. The reduced 2D spectral exposures were sky-subtracted, extracted, stacked, and flux calibrated relative to LTT 1788 using custom IDL routines that make use of the XIDL package\footnote{http://www.ucolick.org/\textasciitilde xavier/IDL/}. Velocity measurements were made using the RVSAO\cite{kurtz98} package with the \emph{fabtemp97} template. The final histogram of velocities, along with the biweight estimate of the redshift and velocity dispersion, are shown in Figure \ref{velhist}.

\subsection{Near-Infrared Spectroscopy}
\begin{figure*}[t]
\begin{center}
\includegraphics[width=0.6\textwidth]{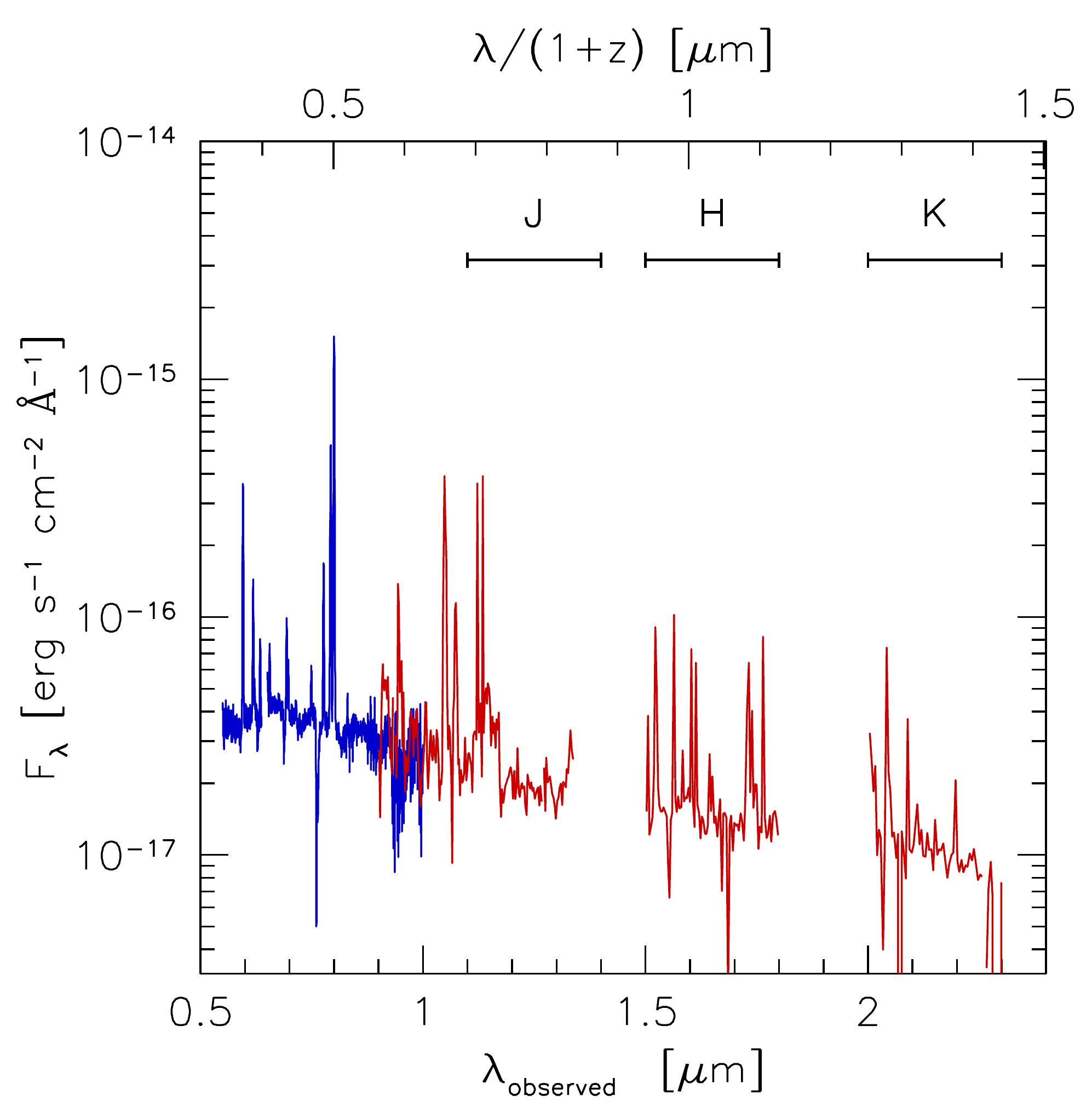}
\caption{IMACS optical (blue) and FIRE infrared (red) spectra of the central galaxy in SPT-CLJ2344-4243. The IMACS spectrum has been scaled down to account for the significantly shorter and narrower FIRE slit.}
\label{firespec}
\end{center}
\end{figure*}

A near-infrared spectrum of SPT-CLJ2344-4243 was obtained with the Folded-port Infrared Echellette (FIRE) spectrograph at the Magellan Baade telescope in late January, 2012. FIRE delivers $R=6000$ spectra between $0.82-2.5$ microns in a single-object, cross-dispersed setup\cite{sim10}. A 10-minute on-target exposure with $\sim$0.9$^{\prime\prime}$ seeing produced the data presented in Figure \ref{firespec}.

For the extraction of point sources, FIRE's reduction pipeline (FIREHOSE) nominally creates a 2-dimensional sky model derived from the portions of the slit that are not illuminated by the source. This way, the sky flux is measured simultaneously with the object flux. Since the spatial extent of SPT-CLJ2344-4243 fills FIRE's $6^{\prime\prime}$-long echelle slit, a separate 10-minute sky exposure had to be obtained and subtracted from the object frame prior to extraction of the spectrum. The variability in the sky over the fifteen minutes between the object and sky exposures can introduce uncertainties into the extracted object spectrum, particularly near hydroxyl (OH) lines. Fortunately, the OH lines subtract with few residuals in the vicinity of the H$\alpha$ emission.

Flux calibration was performed by obtaining the spectrum of an A0V star with an airmass, angular position, and observing time as close to the target as possible. Telluric absorption was corrected\cite{vacca03} via the xtellcor procedure released with the spextool pipeline\cite{cushing04}.

Since FIRE operates in quasi-Littrow mode, the spectral orders are significantly curved and tilted with respect to the detector's pixel basis. Hence, producing a spatial by spectral image of SPT-CLJ2344-4243 required a separate boxcar extraction of the object spectrum for each spatial position. A boxcar width of $0.2^{\prime\prime}$ was selected to match the spatial scale of IMACS. Each of the spectral strips produced from these extractions were telluric-corrected with the observation of an AOV standard.

\section{Optical Line Ratios}

\begin{figure*}[p]
\begin{center}
\includegraphics[width=0.95\textwidth]{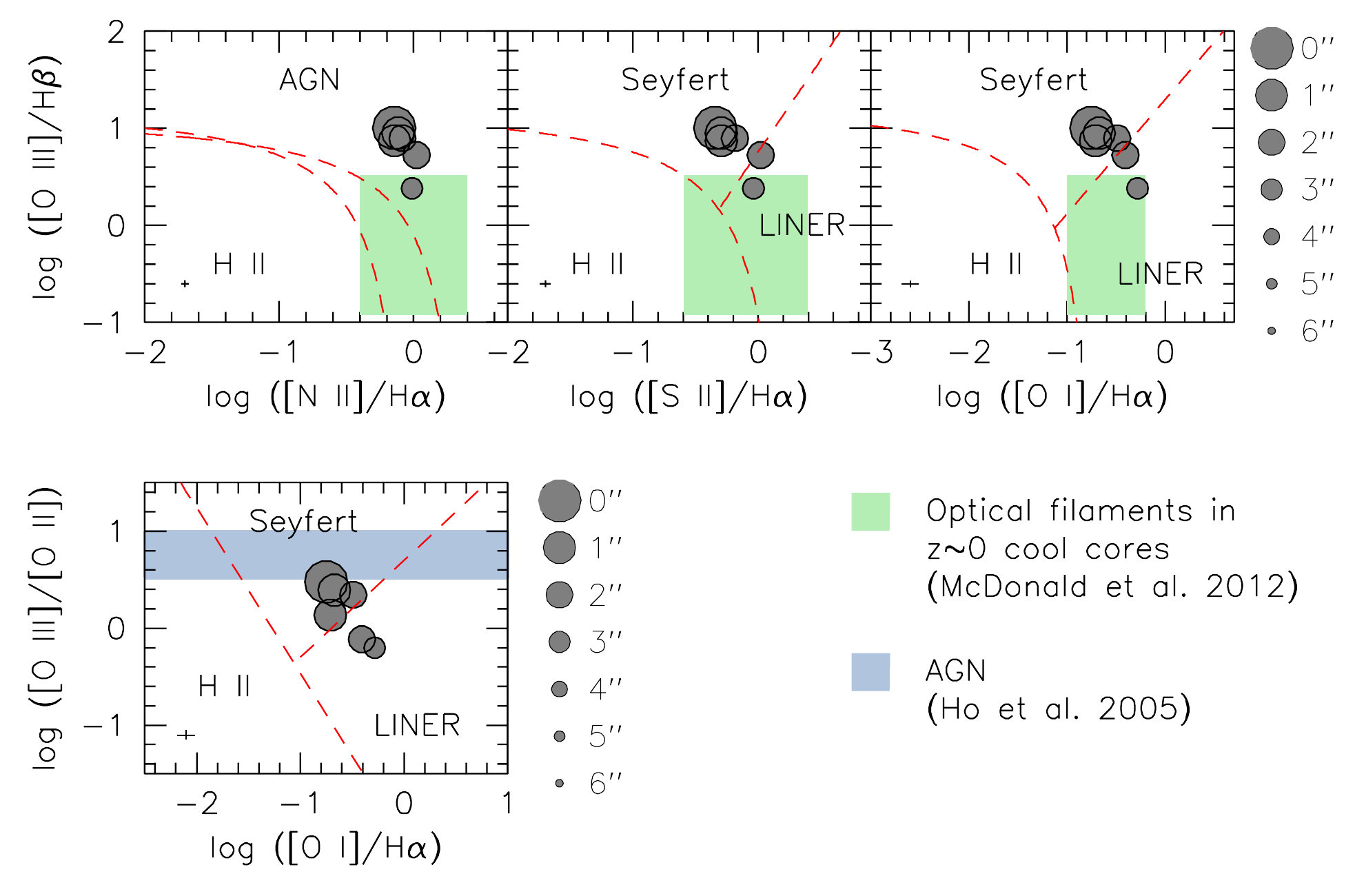}
\caption{Optical line ratio diagnostic plots for the central galaxy in SPT-CLJ2344-4243. The upper panels show [N II] $\lambda$6583/H$\alpha$, [S II] $\lambda\lambda$6716,6731/H$\alpha$, and [O I] $\lambda$6300/H$\alpha$ as a function of [O III] $\lambda$5007/H$\beta$ for a variety of radial bins (in both positive and negative directions along the slit). While the nucleus of this galaxy shows signatures of a strong AGN, the line ratios at large radii resemble more closely the LINER-like spectra of star-forming filaments in z$\sim$0 cool core clusters\cite{mcdonald12a}. In the lower left panel, the [O III] $\lambda$5007/[O II] $\lambda\lambda$3726,3729 ratio is compared to the [O I] $\lambda$6300/H$\alpha$ ratio. Even in the central bin, the [O III]/[O II] ratio is less than expected for AGN\cite{ho05}, suggesting the presence of an additional starburst component. }
\label{lineratios}
\end{center}
\end{figure*}

In order to understand the origin of the bright emission lines
observed in SPT-CLJ2344-4243, we appeal to a variety of optical
emission line ratios which are traditionally used to separate HII
regions from AGN of different
types\cite{veilleux87,kewley06,trouille11}.  In particular, we can
differentiate using line ratios between those typically observed in
Seyfert galaxies and low-ionization nuclear emission-line regions
(LINERs). In the upper panels of Figure \ref{lineratios} we show the
[O III] $\lambda$5007/H$\beta$ ratio as a function of [N II]
$\lambda$6583/H$\alpha$, [S II] $\lambda\lambda$6716,6731/H$\alpha$,
and [O I] $\lambda$6300/H$\alpha$ for the inner 3$^{\prime\prime}$. We
are limited to small radii in these panels due to the small width
($\sim$6$^{\prime\prime}$) of the FIRE echelle slit. These line ratios
confirm that there is a strong AGN in the central region of
SPT-CLJ2344-4243, consistent with the hard X-ray and strong radio
emission. However, the [O III]/H$\beta$ ratio declines quickly with
radius, such that at a distance of 3$^{\prime\prime}$ (20 kpc) from
the nucleus of SPT-CLJ2344-4243, the optical line ratios are similar
to those measured in the cool filaments of z$\sim$0 cool core
clusters\cite{mcdonald12a}.  In the local Universe, these filaments
appear to be ionized by a combination of young stars and shocks,
producing LINER-like line ratios\cite{mcdonald12a}. In the lower left
panel of Figure \ref{lineratios}, we plot [O III] $\lambda$5007/[O II]
$\lambda\lambda$3726,3729 versus [O I] $\lambda$6300/H$\alpha$,
following Kewley \etal (2006). Again, we find that the nucleus has
Seyfert-like line ratios, while the extended emission is
LINER-like. At all radii, the [O III]/[O II] ratio is below the range
quoted in Ho \etal (2005) for AGN (0.1-0.3), further demonstrating
that the ionization in the central region comes from a mix of AGN and
starburst.

We note that the near-IR spectra have a much smaller slit width
(0.6$^{\prime\prime}$ than the optical spectra (1.2$^{\prime\prime}$),
which will result in a slight bias to Figure \ref{lineratios}. Since
the ``nuclear'' points from the optical spectra will be sampling large
radii than those from the near-IR spectra, we expect the [O
  III]/H$\beta$ ratio to be biased low at small radii. Using a narrow
slit may yield an [O III]/[O II] more consistent with pure AGN (lower
left panel of Figure \ref{lineratios}. We stress that the overall
trend, with line ratios resembling AGN at small radii and low-$z$
emission-line nebulae in cool cores at large radii, would not be
caused by the difference in slit widths.


\section{Estimating the Star Formation Rate}

The broad wavelength coverage of the central galaxy in SPT-CLJ2344-4243 allows for a careful estimate of the star formation rate (SFR) using a variety of different techniques. In order to properly constrain the SFR, we must estimate i) the intrinsic reddening, ii) the contribution to the UV continuum and emission line flux from the central AGN, and iii) the amount of line emission missed by the narrow slit. As an initial estimate, we consider the lower limit case which involves masking the central PSF (removing both the AGN and the central starburst component in the central 1$^{\prime\prime}$), assuming zero intrinsic reddening, and assuming that the slit contains 100\% of the extended emission-line flux. Under these assumptions, we calculate a \emph{lower limit} on the star formation rate from the [O II] line of SFR$_{[O II]}$ $>$ 116 M$_{\odot}$ yr$^{-1}$. We note that, despite being a lower limit, this is still higher than any of the lower-redshift, strongly-cooling clusters from Table 1.

In order to estimate the amount of intrinsic reddening, we use the relative fluxes of the H$\beta$, H$\gamma$, and H$\delta$ lines. Assuming case B recombination, we estimate average intrinsic reddening values of E(B-V)$_{\gamma,\beta}$ = 0.35 and E(B-V)$_{\delta,\beta}$ = 0.33. Averaging these two values yields our final estimate of the intrinsic reddening of E(B-V) = 0.34. We correct all of our measured line fluxes for this amount of reddening assuming a dust screen model. If, instead, we assumed a mixed dust/emission scenario, the result would imply a larger extinction correction and, thus, a larger inferred star formation rate. Since the emission line gas may be tracing the youngest, most massive stars (e.g., in low-z starburst galaxies), we expect them to be the most highly enshrouded in dust. However, the stars which are producing the strong UV and optical continuum are slightly older and more dispersed than this very young population. This leads to roughly a factor of 50\% less extinction in the stellar continuum at H$\alpha$ than in the emission line itself\cite{calzetti94}. Thus, we correct all broadband optical and UV fluxes assuming a more conservative intrinsic reddening of E(B-V) = 0.17.

Table S.1 lists the SFRs inferred from extinction-corrected H$\alpha$, H$\beta$, and [O II] emission lines and far-UV, near-UV, optical, and mid--far IR continuum. We assume a Salpeter IMF\cite{salpeter55} for all estimates, using the conversions M$_{Cha}$ = 0.55$\times$M$_{Salp}$ and M$_{Krou}$ = 0.62$\times$M$_{Salp}$. Due to the smaller slit used for the near-IR spectrum, the H$\alpha$-determined SFR is $\sim$50\% less than the SFR estimated from the [O II] and H$\beta$ emission lines. Excluding this low estimate, we find that $<SFR>_{emission~line}$ $\sim$ 730 M$_{\odot}$ yr$^{-1}$, and $<SFR>_{continuum}$ $\sim$ 1120 M$_{\odot}$ yr$^{-1}$. This implies that the slit contains $\sim$65\% of the total flux from star-forming regions. We find remarkable correspondence between the star formation rates inferred from the rest-frame far-UV (GALEX near-UV), near-UV (IMACS $g$-band), 4000\AA\ break, and WISE 24$\mu$m emission, suggesting that calibration errors are negligible. 










\begin{table}[p]
\centering
{\footnotesize
\begin{tabular}{c c c c c c c}
\hline\hline
Method & SFR Estimate & $f_{AGN}$ & Coverage & Reference & Notes\\
 & [M$_{\odot}$ yr$^{-1}$] & & & & \\
\hline
H$\alpha$ &  426 $\pm$ 20 & 0\% & Slit & K98& Small aperture\\
H$\beta$ &  606 $\pm$ 40 & 0\% & Slit & K98  & \\
$[$O II$]$ & 841 $\pm$ 75  & 0\% & Slit & K04  & \\

\\
FUV & 1401 $\pm$ 468 & 0\% & Total & RG02 & Implicit Extinction\\
FUV & 1126 $\pm$ 105 & 0\% & Total & K98,C94  & 50\% H$\gamma$/H$\beta$ Extinction\\
NUV & 1132 $\pm$ 105 & 0\% & Total & K98,C94  & 50\% H$\gamma$/H$\beta$ Extinction\\
D$_{4000}$ & 1960 $\pm$ 726   & 0\% & Total & B04 & M$_{*,z}$ = 3$\times$10$^{12}$ M$_{\odot}$ \\
24$\mu$m & 1655 $\pm$ 711 & 0\% & Total & R09 &  \\
Far IR & 1642 $\pm$ 190 & 0\% & Total & K98  & \\

\\
Extended $[$O II$]$ & 771 $\pm$ 262 & 40 $\pm$ 20\% & Total$^*$ & K04 &  Inner $\sim$1$^{\prime\prime}$ masked\\
$[$O II$]$ + $[$O III$]$ & 799 $\pm$ 232 & 37 $\pm$ 18\% & Total$^*$ & K04, H05 & $[$O II$]$/$[$O III$]$$_{AGN}$ = 0.2 $\pm$ 0.1\\
FUV + X-ray & 612 $\pm$ 207& 45 $\pm$ 17\% & Total & K98, C94, E94 & $\alpha_{UV-X}$ = 1.325 $\pm$ 0.075\\
FIR + X-ray & 774 $\pm$ 493 & 51 $\pm$ 30\% & Total & K98, M11  & \\

\hline
Average: & 739 $\pm$ 160 & $\sim$40-50\% & Total$^*$ & \multicolumn{2}{r}{$^*${\footnotesize: Assumes that slit contains $\sim$65\% of total emission}}\\
\end{tabular}
\caption{Star formation rate estimates for the central galaxy in SPT-CLJ2344-4243 from a variety of methods spanning a large range in wavelength. The average SFR is based on the bottom four estimates, which are corrected for intrinsic extinction, covering fraction, and have had the contribution from the central AGN removed. Uncertainty estimates include contributions from measurement error (typically negligible, with the exception of the far-IR), as well as uncertainty in the intrinsic reddening and the AGN fraction, the latter being the dominant source of error. References are: B04 = Brinchmann \etal (2004)\nocite{brinchmann04}, C94 = Calzetti \etal (1994)\nocite{calzetti94}, E94 = Elvis \etal (1994)\nocite{elvis94}, H05 = Ho \etal (2005)\nocite{ho05}, K98 = Kennicut (1998)\nocite{kennicutt98}, K04 = Kewley \etal (2004)\nocite{kewley04}, M11 = Mullaney \etal (2011)\nocite{mullaney11}, R09 = Rieke \etal (2009)\nocite{rieke09}, RG02 = Rosa-Gonz\'{a}lez (2002)\nocite{rosa-gonzalez02}.}
}
\label{sfrtable}
\end{table}

To properly remove the contribution from the central AGN to the inferred SFR, we apply four distinct corrections. The first correction, which is the most conservative, involves masking the central $\sim$1$^{\prime\prime}$ of the IMACS spectrum. This correction removes both the central AGN and starburst contribution to the total [O II] flux. The second correction, from Ho \etal (2005), assumes that [O II]/[O III]$_{AGN}$ = 0.2 $\pm$ 0.1 (see Figure \ref{lineratios}), and that any additional [O II] emission is from ongoing star formation. Our third correction assumes a constant X-ray--UV ratio for the AGN, from Elvis \etal (1994). We model the X-ray spectrum from the central 1.5$^{\prime\prime}$ with a combined plasma and absorbed powerlaw model, allowing an estimate of the unabsorbed X-ray flux from the central AGN, which is then converted to an estimated contribution to the total UV flux. Finally, the last correction employs a scaling relation between the hard X-ray and the total IR luminosity of AGN, from Mullaney \etal (2011), which provides an estimate of AGN contamination in the measured IR luminosity. The results of these four independent methods (Table S.1) yield a range of AGN contamination from $\sim$40-50\% over a broad range in wavelength. 

Combining the above estimates of the intrinsic reddening, the filling-factor of the slit, and the amount of contamination from the central AGN, we find an average SFR of 739 $\pm$ 160 M$_{\odot}$ yr$^{-1}$. The uncertainty quoted here is a combination of measurement error, along with uncertainty in the amount of intrinsic extinction and the amount of AGN contamination. The relatively small scatter in the SFR estimates from a variety of different methods and wavelengths suggests that this result is robust. However, there are several uncertainties which could conspire to bias this number high. While we expect shocks to contribute to the emission-line spectrum\cite{mcdonald12a}, the fact that the IR- and UV-derived SFRs are higher than those from emission line suggests that this contribution is smaller than our uncertainty in the extinction correction. We have assumed case B recombination in estimating the amount of reddening, which may not reflect the true nature of this system. We note, however, that this correction would only change by $\sim$2-3\% if the reality is closer to case A. Our extinction correction most strongly effects the far-UV data, resulting in a factor of $\sim$3.6 correction to the far-UV luminosity. However, we stress that this significant extinction correction is physically motivated, based on the H$\gamma$/H$\beta$ and H$\delta$/H$\beta$ ratios, the slope of the UV-optical SED, the strong extinction measured in the X-ray, and the strong mid-IR flux. Further, the fact that the empirical SFR estimate from Rosa-Gonzalez \etal (2002) \emph{over estimates} the SFR compared to our reddening-corrected estimates suggests that our assumption of E(B-V) = 0.17 in the stellar continuum is \emph{below average} for systems with similar UV luminosity. In order to improve upon the accuracy of this estimate, we require high spatial resolution UV imaging, which is the least sensitive to shocks and would allow us to spatially model and subtract the nuclear component.

The flux in the SPT 1.4 mm data at the position of the BCG is consistent with zero, and a 3 sigma upper limit can be placed at 20.4 mJy. By fitting a range of SEDs to the available data (Figure 3), we estimate the median 1.4 mm flux to be 0.5 mJy, with a 68\% confidence interval of 0.2--1.4 mJy, and the median 2.0 mm flux to be 0.1 mJy with a 68\% confidence interval of 0.05--0.38 mJy. This level of flux contribution at 2.0 mm would have a negligible effect on SZ flux estimate. We also estimate the contribution to the total X-ray luminosity from the starburst based on scaling relations\cite{ranalli03} and find that a starburst of $\sim$800 M$_{\odot}$ yr$^{-1}$ should have an X-ray luminosity of L$_{2-10keV}$ = 4$\times$10$^{42}$ erg s$^{-1}$, which is less than 0.05\% of the total luminosity presented in Table 1. This suggests that the starburst is contributing a negligible fraction of the total X-ray luminosity in the galaxy cluster core.

\section{Properties of the Central AGN}
The central galaxy in SPT-CLJ2344-4243 appears to host both a strong AGN and a vigorous starburst, both heavily obscured by dust. We summarize the properties of the central AGN in Table S.2. As was mentioned previously, the central point source has high [O III]/H$\beta$ and [O III]/[O II], suggesting a strong Seyfert-like AGN. Consistent with low-redshift cool core clusters, the central galaxy is radio loud ($\nu$L$_{\nu}$ = 10$^{42}$ erg s$^{-1}$)\cite{mauch03}, with a radio luminosity that is similar to those of BCGs in low-z cool cores (e.g., Abell0780, Abell2052), despite orders of magnitude more cooling. This relatively low radio luminosity compared to the cooling luminosity may explain why this cluster appears to be forming stars at such a large fraction of the classical cooling rate. The hard X-ray luminosity of the AGN (L$_{2-10keV}$) suggests an accretion rate of $\frac{dM}{dt}$ = $\frac{L_{bol}}{\eta c^2}$ = 58 M$_{\odot}$ yr$^{-1}$, assuming a black hole accretion efficiency of $\eta$ = 0.1 and a bolometric correction factor (L$_{bol}$/L$_{2-10\rm{keV}}$) of 110\cite{marconi04}. This accretion rate represents a small fraction ($<$2\%) of the total cooling rate, suggesting that either feedback is preventing the cooling flow from efficiently accreting onto the central black hole, or the bulk of the cool material is in some unobserved phase (e.g., cold molecular gas). 






\begin{table}[tb]
{\small
\centering
\begin{tabular}{l l r}
\hline
$\nu$L$_{\nu, radio}$ & 10$^{42}$ erg s$^{-1}$ & (843MHz SUMMS survey)\\
L$_{2-10keV}$ & 3$\times$10$^{45}$ erg s$^{-1}$ & (unobscured X-ray luminosity)\\
L$_{IR}$ & $\sim$1.5$\times$10$^{46}$ erg s$^{-1}$  & (assume AGN is $\sim$40\% of total IR flux)\\
\.M$_{acc}$ & 58 M$_{\odot}$ yr$^{-1}$ & (assume L$_{bol}$/L$_X$ = 110$^{\dagger}$, $\eta_{acc}$ = 0.1)\\
A$_{V,AGN}$ & 1.5 mag & (Balmer extinction)\\
n$_{H,AGN}$ & 39$\times$10$^{22}$ cm$^{-2}$& (X-ray absorption)\\
\hline
$^{\dagger}$: Marconi \etal (2004)\nocite{marconi04}
\end{tabular}
\caption{Properties of the central AGN in SPT-CLJ2344-4243.}
}
\label{agntable}
\end{table}

Based on scaling relations\cite{bennert11} between the total spheroid stellar mass (M$_*$, assuming the spheroid luminosity is equal to the total luminosity) and black hole mass (M$_{BH}$), we naively estimate the supermassive black hole in the central galaxy to have a mass of M$_{BH}$ $\sim$ 1.8$^{+2.5}_{-1.2}$ $\times$10$^{10}$ M$_{\odot}$, which is at the high end of masses seen in the cores of massive galaxy clusters. This high mass would imply that the aforementioned accretion rate of 58 M$_{\odot}$ yr$^{-1}$ is $\sim$15\% of the Eddington rate. 

We stress that both the accretion rate and the black hole mass are highly uncertain, due to the order of magnitude uncertainty in the bolometric correction\cite{ranjan07} and black hole mass\cite{bennert11}. The bolometric correction factor of F$_{bol}$/F$_{2-10keV}$ is likely an upper limit, suggesting that \.M$_{acc}$ $<$ 0.02\.M$_{cool}$.

\section*{References}


\section*{References}

\begin{thebibliography}{10}
\expandafter\ifx\csname url\endcsname\relax
  \def\url#1{\texttt{#1}}\fi
\expandafter\ifx\csname urlprefix\endcsname\relax\def\urlprefix{URL }\fi
\providecommand{\bibinfo}[2]{#2}
\providecommand{\eprint}[2][]{\url{#2}}

\bibitem{lea73}
\bibinfo{author}{{Lea}, S.~M.}, \bibinfo{author}{{Silk}, J.},
  \bibinfo{author}{{Kellogg}, E.} \& \bibinfo{author}{{Murray}, S.}
\newblock \bibinfo{title}{{Thermal-Bremsstrahlung Interpretation of Cluster
  X-Ray Sources}}.
\newblock \emph{\bibinfo{journal}{\apjl}} \textbf{\bibinfo{volume}{184}},
  \bibinfo{pages}{L105} (\bibinfo{year}{1973}).

\bibitem{cowie77}
\bibinfo{author}{{Cowie}, L.~L.} \& \bibinfo{author}{{Binney}, J.}
\newblock \bibinfo{title}{{Radiative regulation of gas flow within clusters of
  galaxies - A model for cluster X-ray sources}}.
\newblock \emph{\bibinfo{journal}{\apj}} \textbf{\bibinfo{volume}{215}},
  \bibinfo{pages}{723--732} (\bibinfo{year}{1977}).

\bibitem{fabian77}
\bibinfo{author}{{Fabian}, A.~C.} \& \bibinfo{author}{{Nulsen}, P.~E.~J.}
\newblock \bibinfo{title}{{Subsonic accretion of cooling gas in clusters of
  galaxies}}.
\newblock \emph{\bibinfo{journal}{\mnras}} \textbf{\bibinfo{volume}{180}},
  \bibinfo{pages}{479--484} (\bibinfo{year}{1977}).

\bibitem{odea08}
\bibinfo{author}{{O'Dea}, C.~P.} \emph{et~al.}
\newblock \bibinfo{title}{{An Infrared Survey of Brightest Cluster Galaxies.
  II. Why are Some Brightest Cluster Galaxies Forming Stars?}}
\newblock \emph{\bibinfo{journal}{\apj}} \textbf{\bibinfo{volume}{681}},
  \bibinfo{pages}{1035--1045} (\bibinfo{year}{2008}).
\newblock \eprint{0803.1772}.

\bibitem{mcdonald11b}
\bibinfo{author}{{McDonald}, M.}, \bibinfo{author}{{Veilleux}, S.},
  \bibinfo{author}{{Rupke}, D.~S.~N.}, \bibinfo{author}{{Mushotzky}, R.} \&
  \bibinfo{author}{{Reynolds}, C.}
\newblock \bibinfo{title}{{Star Formation Efficiency in the Cool Cores of
  Galaxy Clusters}}.
\newblock \emph{\bibinfo{journal}{\apj}} \textbf{\bibinfo{volume}{734}},
  \bibinfo{pages}{95} (\bibinfo{year}{2011}).
\newblock \eprint{1104.0665}.

\bibitem{edge01}
\bibinfo{author}{{Edge}, A.~C.}
\newblock \bibinfo{title}{{The detection of molecular gas in the central
  galaxies of cooling flow clusters}}.
\newblock \emph{\bibinfo{journal}{\mnras}} \textbf{\bibinfo{volume}{328}},
  \bibinfo{pages}{762--782} (\bibinfo{year}{2001}).
\newblock \eprint{arXiv:astro-ph/0106225}.

\bibitem{mcnamara07}
\bibinfo{author}{{McNamara}, B.~R.} \& \bibinfo{author}{{Nulsen}, P.~E.~J.}
\newblock \bibinfo{title}{{Heating Hot Atmospheres with Active Galactic
  Nuclei}}.
\newblock \emph{\bibinfo{journal}{\araa}} \textbf{\bibinfo{volume}{45}},
  \bibinfo{pages}{117--175} (\bibinfo{year}{2007}).
\newblock \eprint{0709.2152}.

\bibitem{fabian12}
\bibinfo{author}{{Fabian}, A.~C.}
\newblock \bibinfo{title}{{Observational Evidence of AGN Feedback}}.
\newblock \emph{\bibinfo{journal}{ArXiv e-prints}}  (\bibinfo{year}{2012}).
\newblock \eprint{1204.4114}.

\bibitem{mathews09}
\bibinfo{author}{{Mathews}, W.~G.}
\newblock \bibinfo{title}{{Stopping Cooling Flows with Cosmic-Ray Feedback}}.
\newblock \emph{\bibinfo{journal}{\apjl}} \textbf{\bibinfo{volume}{695}},
  \bibinfo{pages}{L49--L52} (\bibinfo{year}{2009}).
\newblock \eprint{0903.1135}.

\bibitem{gomez02}
\bibinfo{author}{{G{\'o}mez}, P.~L.}, \bibinfo{author}{{Loken}, C.},
  \bibinfo{author}{{Roettiger}, K.} \& \bibinfo{author}{{Burns}, J.~O.}
\newblock \bibinfo{title}{{Do Cooling Flows Survive Cluster Mergers?}}
\newblock \emph{\bibinfo{journal}{\apj}} \textbf{\bibinfo{volume}{569}},
  \bibinfo{pages}{122--133} (\bibinfo{year}{2002}).

\bibitem{williamson11}
\bibinfo{author}{{Williamson}, R.} \emph{et~al.}
\newblock \bibinfo{title}{{A Sunyaev-Zel'dovich-selected Sample of the Most
  Massive Galaxy Clusters in the 2500 deg$^{2}$ South Pole Telescope Survey}}.
\newblock \emph{\bibinfo{journal}{\apj}} \textbf{\bibinfo{volume}{738}},
  \bibinfo{pages}{139} (\bibinfo{year}{2011}).
\newblock \eprint{1101.1290}.

\bibitem{carlstrom11}
\bibinfo{author}{{Carlstrom}, J.~E.} \emph{et~al.}
\newblock \bibinfo{title}{{The 10 Meter South Pole Telescope}}.
\newblock \emph{\bibinfo{journal}{\pasp}} \textbf{\bibinfo{volume}{123}},
  \bibinfo{pages}{568--581} (\bibinfo{year}{2011}).
\newblock \eprint{0907.4445}.

\bibitem{vikhlinin09a}
\bibinfo{author}{{Vikhlinin}, A.} \emph{et~al.}
\newblock \bibinfo{title}{{Chandra Cluster Cosmology Project. II. Samples and
  X-Ray Data Reduction}}.
\newblock \emph{\bibinfo{journal}{\apj}} \textbf{\bibinfo{volume}{692}},
  \bibinfo{pages}{1033--1059} (\bibinfo{year}{2009}).
\newblock \eprint{0805.2207}.

\bibitem{menanteau11}
\bibinfo{author}{{Menanteau}, F.} \emph{et~al.}
\newblock \bibinfo{title}{{The Atacama Cosmology Telescope: ACT-CL J0102-4915
  ''El Gordo,'' a Massive Merging Cluster at Redshift 0.87}}.
\newblock \emph{\bibinfo{journal}{ArXiv e-prints}}  (\bibinfo{year}{2011}).
\newblock \eprint{1109.0953}.

\bibitem{foley11}
\bibinfo{author}{{Foley}, R.~J.} \emph{et~al.}
\newblock \bibinfo{title}{{Discovery and Cosmological Implications of SPT-CL
  J2106-5844, the Most Massive Known Cluster at z$>$1}}.
\newblock \emph{\bibinfo{journal}{\apj}} \textbf{\bibinfo{volume}{731}},
  \bibinfo{pages}{86} (\bibinfo{year}{2011}).
\newblock \eprint{1101.1286}.

\bibitem{fabian00}
\bibinfo{author}{{Fabian}, A.~C.} \emph{et~al.}
\newblock \bibinfo{title}{{Chandra imaging of the complex X-ray core of the
  Perseus cluster}}.
\newblock \emph{\bibinfo{journal}{\mnras}} \textbf{\bibinfo{volume}{318}},
  \bibinfo{pages}{L65--L68} (\bibinfo{year}{2000}).
\newblock \eprint{arXiv:astro-ph/0007456}.

\bibitem{allen96}
\bibinfo{author}{{Allen}, S.~W.}, \bibinfo{author}{{Fabian}, A.~C.} \&
  \bibinfo{author}{{Kneib}, J.~P.}
\newblock \bibinfo{title}{{A combined X-ray and gravitational lensing study of
  the massive cooling-flow cluster PKS 0745-191}}.
\newblock \emph{\bibinfo{journal}{\mnras}} \textbf{\bibinfo{volume}{279}},
  \bibinfo{pages}{615--635} (\bibinfo{year}{1996}).
\newblock \eprint{arXiv:astro-ph/9506036}.

\bibitem{vikhlinin07}
\bibinfo{author}{{Vikhlinin}, A.} \emph{et~al.}
\newblock \bibinfo{title}{{Lack of Cooling Flow Clusters at z $>$ 0.5}}.
\newblock In \bibinfo{editor}{{B{\"o}hringer}, H.}, \bibinfo{editor}{{Pratt},
  G.~W.}, \bibinfo{editor}{{Finoguenov}, A.} \& \bibinfo{editor}{{Schuecker},
  P.} (eds.) \emph{\bibinfo{booktitle}{Heating versus Cooling in Galaxies and
  Clusters of Galaxies}}, \bibinfo{pages}{48} (\bibinfo{year}{2007}).
\newblock \eprint{arXiv:astro-ph/0611438}.

\bibitem{santos08}
\bibinfo{author}{{Santos}, J.~S.} \emph{et~al.}
\newblock \bibinfo{title}{{Searching for cool core clusters at high redshift}}.
\newblock \emph{\bibinfo{journal}{\aap}} \textbf{\bibinfo{volume}{483}},
  \bibinfo{pages}{35--47} (\bibinfo{year}{2008}).
\newblock \eprint{0802.1445}.

\bibitem{mcdonald11c}
\bibinfo{author}{{McDonald}, M.}
\newblock \bibinfo{title}{{Optical Line Emission in Brightest Cluster Galaxies
  at 0 $<$ z $<$ 0.6: Evidence for a Lack of Strong Cool Cores 3.5 Gyr Ago?}}
\newblock \emph{\bibinfo{journal}{\apjl}} \textbf{\bibinfo{volume}{742}},
  \bibinfo{pages}{L35} (\bibinfo{year}{2011}).
\newblock \eprint{1110.5904}.

\bibitem{hu85}
\bibinfo{author}{{Hu}, E.~M.}, \bibinfo{author}{{Cowie}, L.~L.} \&
  \bibinfo{author}{{Wang}, Z.}
\newblock \bibinfo{title}{{Long-slit spectroscopy of gas in the cores of X-ray
  luminous clusters}}.
\newblock \emph{\bibinfo{journal}{\apjs}} \textbf{\bibinfo{volume}{59}},
  \bibinfo{pages}{447--498} (\bibinfo{year}{1985}).

\bibitem{heckman89}
\bibinfo{author}{{Heckman}, T.~M.}, \bibinfo{author}{{Baum}, S.~A.},
  \bibinfo{author}{{van Breugel}, W.~J.~M.} \& \bibinfo{author}{{McCarthy}, P.}
\newblock \bibinfo{title}{{Dynamical, physical, and chemical properties of
  emission-line nebulae in cooling flows}}.
\newblock \emph{\bibinfo{journal}{\apj}} \textbf{\bibinfo{volume}{338}},
  \bibinfo{pages}{48--77} (\bibinfo{year}{1989}).

\bibitem{mcdonald10}
\bibinfo{author}{{McDonald}, M.}, \bibinfo{author}{{Veilleux}, S.},
  \bibinfo{author}{{Rupke}, D.~S.~N.} \& \bibinfo{author}{{Mushotzky}, R.}
\newblock \bibinfo{title}{{On the Origin of the Extended H{$\alpha$} Filaments
  in Cooling Flow Clusters}}.
\newblock \emph{\bibinfo{journal}{\apj}} \textbf{\bibinfo{volume}{721}},
  \bibinfo{pages}{1262--1283} (\bibinfo{year}{2010}).
\newblock \eprint{1008.0392}.

\bibitem{conselice01}
\bibinfo{author}{{Conselice}, C.~J.}, \bibinfo{author}{{Gallagher}, J.~S., III}
  \& \bibinfo{author}{{Wyse}, R.~F.~G.}
\newblock \bibinfo{title}{{On the Nature of the NGC 1275 System}}.
\newblock \emph{\bibinfo{journal}{\aj}} \textbf{\bibinfo{volume}{122}},
  \bibinfo{pages}{2281--2300} (\bibinfo{year}{2001}).
\newblock \eprint{arXiv:astro-ph/0108019}.

\bibitem{sparks96}
\bibinfo{author}{{Sparks}, W.~B.}, \bibinfo{author}{{Biretta}, J.~A.} \&
  \bibinfo{author}{{Macchetto}, F.}
\newblock \bibinfo{title}{{The Jet of M87 at Tenth-Arcsecond Resolution:
  Optical, Ultraviolet, and Radio Observations}}.
\newblock \emph{\bibinfo{journal}{\apj}} \textbf{\bibinfo{volume}{473}},
  \bibinfo{pages}{254} (\bibinfo{year}{1996}).

\bibitem{mcdonald12a}
\bibinfo{author}{{McDonald}, M.}, \bibinfo{author}{{Veilleux}, S.} \&
  \bibinfo{author}{{Rupke}, D.~S.~N.}
\newblock \bibinfo{title}{{Optical Spectroscopy of H{$\alpha$} Filaments in
  Cool Core Clusters: Kinematics, Reddening, and Sources of Ionization}}.
\newblock \emph{\bibinfo{journal}{\apj}} \textbf{\bibinfo{volume}{746}},
  \bibinfo{pages}{153} (\bibinfo{year}{2012}).
\newblock \eprint{1111.0006}.

\bibitem{mcnamara06}
\bibinfo{author}{{McNamara}, B.~R.} \emph{et~al.}
\newblock \bibinfo{title}{{The Starburst in the Abell 1835 Cluster Central
  Galaxy: A Case Study of Galaxy Formation Regulated by an Outburst from a
  Supermassive Black Hole}}.
\newblock \emph{\bibinfo{journal}{\apj}} \textbf{\bibinfo{volume}{648}},
  \bibinfo{pages}{164--175} (\bibinfo{year}{2006}).
\newblock \eprint{arXiv:astro-ph/0604044}.

\bibitem{allen00}
\bibinfo{author}{{Allen}, S.~W.}
\newblock \bibinfo{title}{{The properties of cooling flows in X-ray luminous
  clusters of galaxies}}.
\newblock \emph{\bibinfo{journal}{\mnras}} \textbf{\bibinfo{volume}{315}},
  \bibinfo{pages}{269--295} (\bibinfo{year}{2000}).
\newblock \eprint{arXiv:astro-ph/0002506}.

\bibitem{gitti04}
\bibinfo{author}{{Gitti}, M.} \& \bibinfo{author}{{Schindler}, S.}
\newblock \bibinfo{title}{{XMM-Newton observation of the most X-ray-luminous
  galaxy cluster RX J1347.5-1145}}.
\newblock \emph{\bibinfo{journal}{\aap}} \textbf{\bibinfo{volume}{427}},
  \bibinfo{pages}{L9--L12} (\bibinfo{year}{2004}).
\newblock \eprint{arXiv:astro-ph/0409627}.

\bibitem{polletta07}
\bibinfo{author}{{Polletta}, M.} \emph{et~al.}
\newblock \bibinfo{title}{{Spectral Energy Distributions of Hard X-Ray Selected
  Active Galactic Nuclei in the XMM-Newton Medium Deep Survey}}.
\newblock \emph{\bibinfo{journal}{\apj}} \textbf{\bibinfo{volume}{663}},
  \bibinfo{pages}{81--102} (\bibinfo{year}{2007}).
\newblock \eprint{arXiv:astro-ph/0703255}.
\end{thebibliography}

\begin{thebibliography}{10}
\expandafter\ifx\csname url\endcsname\relax
  \def\url#1{\texttt{#1}}\fi
\expandafter\ifx\csname urlprefix\endcsname\relax\def\urlprefix{URL }\fi
\providecommand{\bibinfo}[2]{#2}
\providecommand{\eprint}[2][]{\url{#2}}

\setcounter{enumiv}{30}

\bibitem{vikhlinin05}
\bibinfo{author}{{Vikhlinin}, A.} \emph{et~al.}
\newblock \bibinfo{title}{{Chandra Temperature Profiles for a Sample of Nearby
  Relaxed Galaxy Clusters}}.
\newblock \emph{\bibinfo{journal}{\apj}} \textbf{\bibinfo{volume}{628}},
  \bibinfo{pages}{655--672} (\bibinfo{year}{2005}).
\newblock \eprint{arXiv:astro-ph/0412306}.

\bibitem{A11}
\bibinfo{author}{{Andersson}, K.} \emph{et~al.}
\newblock \bibinfo{title}{{X-Ray Properties of the First Sunyaev-Zel'dovich
  Effect Selected Galaxy Cluster Sample from the South Pole Telescope}}.
\newblock \emph{\bibinfo{journal}{\apj}} \textbf{\bibinfo{volume}{738}},
  \bibinfo{pages}{48} (\bibinfo{year}{2011}).
\newblock \eprint{1006.3068}.

\bibitem{vikhlinin09a}
\bibinfo{author}{{Vikhlinin}, A.} \emph{et~al.}
\newblock \bibinfo{title}{{Chandra Cluster Cosmology Project. II. Samples and
  X-Ray Data Reduction}}.
\newblock \emph{\bibinfo{journal}{\apj}} \textbf{\bibinfo{volume}{692}},
  \bibinfo{pages}{1033--1059} (\bibinfo{year}{2009}).
\newblock \eprint{0805.2207}.

\bibitem{duffy08}
\bibinfo{author}{{Duffy}, A.~R.}, \bibinfo{author}{{Schaye}, J.},
  \bibinfo{author}{{Kay}, S.~T.} \& \bibinfo{author}{{Dalla Vecchia}, C.}
\newblock \bibinfo{title}{{Dark matter halo concentrations in the Wilkinson
  Microwave Anisotropy Probe year 5 cosmology}}.
\newblock \emph{\bibinfo{journal}{\mnras}} \textbf{\bibinfo{volume}{390}},
  \bibinfo{pages}{L64--L68} (\bibinfo{year}{2008}).
\newblock \eprint{0804.2486}.

\bibitem{odea08}
\bibinfo{author}{{O'Dea}, C.~P.} \emph{et~al.}
\newblock \bibinfo{title}{{An Infrared Survey of Brightest Cluster Galaxies.
  II. Why are Some Brightest Cluster Galaxies Forming Stars?}}
\newblock \emph{\bibinfo{journal}{\apj}} \textbf{\bibinfo{volume}{681}},
  \bibinfo{pages}{1035--1045} (\bibinfo{year}{2008}).
\newblock \eprint{0803.1772}.

\bibitem{blain03}
\bibinfo{author}{{Blain}, A.~W.}, \bibinfo{author}{{Barnard}, V.~E.} \&
  \bibinfo{author}{{Chapman}, S.~C.}
\newblock \bibinfo{title}{{Submillimetre and far-infrared spectral energy
  distributions of galaxies: the luminosity-temperature relation and
  consequences for photometric redshifts}}.
\newblock \emph{\bibinfo{journal}{\mnras}} \textbf{\bibinfo{volume}{338}},
  \bibinfo{pages}{733--744} (\bibinfo{year}{2003}).
\newblock \eprint{arXiv:astro-ph/0209450}.

\bibitem{lacosmic}
\bibinfo{author}{{van Dokkum}, P.~G.}
\newblock \bibinfo{title}{{Cosmic-Ray Rejection by Laplacian Edge Detection}}.
\newblock \emph{\bibinfo{journal}{\pasp}} \textbf{\bibinfo{volume}{113}},
  \bibinfo{pages}{1420--1427} (\bibinfo{year}{2001}).
\newblock \eprint{arXiv:astro-ph/0108003}.

\bibitem{kurtz98}
\bibinfo{author}{{Kurtz}, M.~J.} \& \bibinfo{author}{{Mink}, D.~J.}
\newblock \bibinfo{title}{{RVSAO 2.0: Digital Redshifts and Radial
  Velocities}}.
\newblock \emph{\bibinfo{journal}{\pasp}} \textbf{\bibinfo{volume}{110}},
  \bibinfo{pages}{934--977} (\bibinfo{year}{1998}).
\newblock \eprint{arXiv:astro-ph/9803252}.

\bibitem{sim10}
\bibinfo{author}{{Simcoe}, R.~A.} \emph{et~al.}
\newblock \bibinfo{title}{{The FIRE infrared spectrometer at Magellan:
  construction and commissioning}}.
\newblock In \emph{\bibinfo{booktitle}{Society of Photo-Optical Instrumentation
  Engineers (SPIE) Conference Series}}, vol. \bibinfo{volume}{7735} of
  \emph{\bibinfo{series}{Society of Photo-Optical Instrumentation Engineers
  (SPIE) Conference Series}} (\bibinfo{year}{2010}).

\bibitem{vacca03}
\bibinfo{author}{{Vacca}, W.~D.}, \bibinfo{author}{{Cushing}, M.~C.} \&
  \bibinfo{author}{{Rayner}, J.~T.}
\newblock \bibinfo{title}{{A Method of Correcting Near-Infrared Spectra for
  Telluric Absorption}}.
\newblock \emph{\bibinfo{journal}{\pasp}} \textbf{\bibinfo{volume}{115}},
  \bibinfo{pages}{389--409} (\bibinfo{year}{2003}).
\newblock \eprint{arXiv:astro-ph/0211255}.

\bibitem{cushing04}
\bibinfo{author}{{Cushing}, M.~C.}, \bibinfo{author}{{Vacca}, W.~D.} \&
  \bibinfo{author}{{Rayner}, J.~T.}
\newblock \bibinfo{title}{{Spextool: A Spectral Extraction Package for SpeX, a
  0.8-5.5 Micron Cross-Dispersed Spectrograph}}.
\newblock \emph{\bibinfo{journal}{\pasp}} \textbf{\bibinfo{volume}{116}},
  \bibinfo{pages}{362--376} (\bibinfo{year}{2004}).

\bibitem{veilleux87}
\bibinfo{author}{{Veilleux}, S.} \& \bibinfo{author}{{Osterbrock}, D.~E.}
\newblock \bibinfo{title}{{Spectral classification of emission-line galaxies}}.
\newblock \emph{\bibinfo{journal}{\apjs}} \textbf{\bibinfo{volume}{63}},
  \bibinfo{pages}{295--310} (\bibinfo{year}{1987}).

\bibitem{kewley06}
\bibinfo{author}{{Kewley}, L.~J.}, \bibinfo{author}{{Groves}, B.},
  \bibinfo{author}{{Kauffmann}, G.} \& \bibinfo{author}{{Heckman}, T.}
\newblock \bibinfo{title}{{The host galaxies and classification of active
  galactic nuclei}}.
\newblock \emph{\bibinfo{journal}{\mnras}} \textbf{\bibinfo{volume}{372}},
  \bibinfo{pages}{961--976} (\bibinfo{year}{2006}).
\newblock \eprint{arXiv:astro-ph/0605681}.

\bibitem{trouille11}
\bibinfo{author}{{Trouille}, L.}, \bibinfo{author}{{Barger}, A.~J.} \&
  \bibinfo{author}{{Tremonti}, C.}
\newblock \bibinfo{title}{{The OPTX Project. V. Identifying Distant Active
  Galactic Nuclei}}.
\newblock \emph{\bibinfo{journal}{\apj}} \textbf{\bibinfo{volume}{742}},
  \bibinfo{pages}{46} (\bibinfo{year}{2011}).

\bibitem{mcdonald12a}
\bibinfo{author}{{McDonald}, M.}, \bibinfo{author}{{Veilleux}, S.} \&
  \bibinfo{author}{{Rupke}, D.~S.~N.}
\newblock \bibinfo{title}{{Optical Spectroscopy of H{$\alpha$} Filaments in
  Cool Core Clusters: Kinematics, Reddening, and Sources of Ionization}}.
\newblock \emph{\bibinfo{journal}{\apj}} \textbf{\bibinfo{volume}{746}},
  \bibinfo{pages}{153} (\bibinfo{year}{2012}).
\newblock \eprint{1111.0006}.

\bibitem{ho05}
\bibinfo{author}{{Ho}, L.~C.}
\newblock \bibinfo{title}{{[O II] Emission in Quasar Host Galaxies: Evidence
  for a Suppressed Star Formation Efficiency}}.
\newblock \emph{\bibinfo{journal}{\apj}} \textbf{\bibinfo{volume}{629}},
  \bibinfo{pages}{680--685} (\bibinfo{year}{2005}).
\newblock \eprint{arXiv:astro-ph/0504642}.

\bibitem{calzetti94}
\bibinfo{author}{{Calzetti}, D.}, \bibinfo{author}{{Kinney}, A.~L.} \&
  \bibinfo{author}{{Storchi-Bergmann}, T.}
\newblock \bibinfo{title}{{Dust extinction of the stellar continua in starburst
  galaxies: The ultraviolet and optical extinction law}}.
\newblock \emph{\bibinfo{journal}{\apj}} \textbf{\bibinfo{volume}{429}},
  \bibinfo{pages}{582--601} (\bibinfo{year}{1994}).

\bibitem{salpeter55}
\bibinfo{author}{{Salpeter}, E.~E.}
\newblock \bibinfo{title}{{The Luminosity Function and Stellar Evolution.}}
\newblock \emph{\bibinfo{journal}{\apj}} \textbf{\bibinfo{volume}{121}},
  \bibinfo{pages}{161--+} (\bibinfo{year}{1955}).

\bibitem{brinchmann04}
\bibinfo{author}{{Brinchmann}, J.} \emph{et~al.}
\newblock \bibinfo{title}{{The physical properties of star-forming galaxies in
  the low-redshift Universe}}.
\newblock \emph{\bibinfo{journal}{\mnras}} \textbf{\bibinfo{volume}{351}},
  \bibinfo{pages}{1151--1179} (\bibinfo{year}{2004}).
\newblock \eprint{arXiv:astro-ph/0311060}.

\bibitem{elvis94}
\bibinfo{author}{{Elvis}, M.} \emph{et~al.}
\newblock \bibinfo{title}{{Atlas of quasar energy distributions}}.
\newblock \emph{\bibinfo{journal}{\apjs}} \textbf{\bibinfo{volume}{95}},
  \bibinfo{pages}{1--68} (\bibinfo{year}{1994}).

\bibitem{kennicutt98}
\bibinfo{author}{{Kennicutt}, R.~C., Jr.}
\newblock \bibinfo{title}{{Star Formation in Galaxies Along the Hubble
  Sequence}}.
\newblock \emph{\bibinfo{journal}{\araa}} \textbf{\bibinfo{volume}{36}},
  \bibinfo{pages}{189--232} (\bibinfo{year}{1998}).
\newblock \eprint{arXiv:astro-ph/9807187}.

\bibitem{kewley04}
\bibinfo{author}{{Kewley}, L.~J.}, \bibinfo{author}{{Geller}, M.~J.} \&
  \bibinfo{author}{{Jansen}, R.~A.}
\newblock \bibinfo{title}{{[O II] as a Star Formation Rate Indicator}}.
\newblock \emph{\bibinfo{journal}{\aj}} \textbf{\bibinfo{volume}{127}},
  \bibinfo{pages}{2002--2030} (\bibinfo{year}{2004}).
\newblock \eprint{arXiv:astro-ph/0401172}.

\bibitem{mullaney11}
\bibinfo{author}{{Mullaney}, J.~R.}, \bibinfo{author}{{Alexander}, D.~M.},
  \bibinfo{author}{{Goulding}, A.~D.} \& \bibinfo{author}{{Hickox}, R.~C.}
\newblock \bibinfo{title}{{Defining the intrinsic AGN infrared spectral energy
  distribution and measuring its contribution to the infrared output of
  composite galaxies}}.
\newblock \emph{\bibinfo{journal}{\mnras}} \textbf{\bibinfo{volume}{414}},
  \bibinfo{pages}{1082--1110} (\bibinfo{year}{2011}).
\newblock \eprint{1102.1425}.

\bibitem{rieke09}
\bibinfo{author}{{Rieke}, G.~H.} \emph{et~al.}
\newblock \bibinfo{title}{{Determining Star Formation Rates for Infrared
  Galaxies}}.
\newblock \emph{\bibinfo{journal}{\apj}} \textbf{\bibinfo{volume}{692}},
  \bibinfo{pages}{556--573} (\bibinfo{year}{2009}).
\newblock \eprint{0810.4150}.

\bibitem{rosa-gonzalez02}
\bibinfo{author}{{Rosa-Gonz{\'a}lez}, D.}, \bibinfo{author}{{Terlevich}, E.} \&
  \bibinfo{author}{{Terlevich}, R.}
\newblock \bibinfo{title}{{An empirical calibration of star formation rate
  estimators}}.
\newblock \emph{\bibinfo{journal}{\mnras}} \textbf{\bibinfo{volume}{332}},
  \bibinfo{pages}{283--295} (\bibinfo{year}{2002}).
\newblock \eprint{arXiv:astro-ph/0112556}.

\bibitem{ranalli03}
\bibinfo{author}{{Ranalli}, P.}, \bibinfo{author}{{Comastri}, A.} \&
  \bibinfo{author}{{Setti}, G.}
\newblock \bibinfo{title}{{The 2-10 keV luminosity as a Star Formation Rate
  indicator}}.
\newblock \emph{\bibinfo{journal}{\aap}} \textbf{\bibinfo{volume}{399}},
  \bibinfo{pages}{39--50} (\bibinfo{year}{2003}).
\newblock \eprint{arXiv:astro-ph/0211304}.

\bibitem{mauch03}
\bibinfo{author}{{Mauch}, T.} \emph{et~al.}
\newblock \bibinfo{title}{{SUMSS: a wide-field radio imaging survey of the
  southern sky - II. The source catalogue}}.
\newblock \emph{\bibinfo{journal}{\mnras}} \textbf{\bibinfo{volume}{342}},
  \bibinfo{pages}{1117--1130} (\bibinfo{year}{2003}).
\newblock \eprint{arXiv:astro-ph/0303188}.

\bibitem{marconi04}
\bibinfo{author}{{Marconi}, A.} \emph{et~al.}
\newblock \bibinfo{title}{{Local supermassive black holes, relics of active
  galactic nuclei and the X-ray background}}.
\newblock \emph{\bibinfo{journal}{\mnras}} \textbf{\bibinfo{volume}{351}},
  \bibinfo{pages}{169--185} (\bibinfo{year}{2004}).
\newblock \eprint{arXiv:astro-ph/0311619}.

\bibitem{bennert11}
\bibinfo{author}{{Bennert}, V.~N.}, \bibinfo{author}{{Auger}, M.~W.},
  \bibinfo{author}{{Treu}, T.}, \bibinfo{author}{{Woo}, J.-H.} \&
  \bibinfo{author}{{Malkan}, M.~A.}
\newblock \bibinfo{title}{{A Local Baseline of the Black Hole Mass Scaling
  Relations for Active Galaxies. I. Methodology and Results of Pilot Study}}.
\newblock \emph{\bibinfo{journal}{\apj}} \textbf{\bibinfo{volume}{726}},
  \bibinfo{pages}{59} (\bibinfo{year}{2011}).
\newblock \eprint{1008.4602}.

\bibitem{ranjan07}
\bibinfo{author}{{Vasudevan}, R.~V.} \& \bibinfo{author}{{Fabian}, A.~C.}
\newblock \bibinfo{title}{{Piecing together the X-ray background: bolometric
  corrections for active galactic nuclei}}.
\newblock \emph{\bibinfo{journal}{\mnras}} \textbf{\bibinfo{volume}{381}},
  \bibinfo{pages}{1235--1251} (\bibinfo{year}{2007}).
\newblock \eprint{0708.4308}.

\end{thebibliography}

\end{document}